\documentclass[fleqn,usenatbib]{mnras}

\usepackage{lineno}

\usepackage{newtxtext,newtxmath}

\usepackage[T1]{fontenc}

\DeclareRobustCommand{\VAN}[3]{#2}
\let\VANthebibliography\thebibliography
\def\thebibliography{\DeclareRobustCommand{\VAN}[3]{##3}\VANthebibliography}

\usepackage{graphicx}	
\usepackage{amsmath}	

\graphicspath{{figs/}}
\renewcommand{\arraystretch}{1.1}

\usepackage[dvipsnames]{xcolor}
\usepackage{multirow}

\newcommand{\Lya}{\ensuremath{\text{Ly}\,\alpha}}

\newcommand{\xHI}{\ensuremath{\overline{x}_\textsc{hi}}}
\newcommand{\Muv}{\ensuremath{M_\textsc{uv}}}
\newcommand{\Msun}{\ensuremath{\text{M}_\odot}}
\newcommand{\TIGM}{\ensuremath{\mathcal{T}_\textsc{igm}}}
\newcommand{\EW}{\ensuremath{\text{EW}}}
\newcommand{\zsys}{\ensuremath{z_\text{sys}}}
\newcommand{\beagle}{\texttt{BEAGLE}}

\title[$z = 8.7$ \Lya{} emission]{Deep \textit{JWST} spectroscopy of galaxies in a candidate ionized bubble at $z = 8.7$: probing reionization at pMpc scales with \Lya{} emission}

\author[Whitler et al.]{Lily Whitler,$^{1,2,3}$\thanks{Corresponding author: \href{mailto:lw851@cam.ac.uk}{lw851@cam.ac.uk}}
Daniel P. Stark,$^{4}$
Charlotte A. Mason,$^{5,6}$
Mengtao Tang,$^{3}$
Zuyi Chen,$^{3,5,6}$
\newauthor
Ting-Yi Lu,$^{7}$
Gonzalo Prieto-Lyon,$^{5,6}$
and Anne Hutter$^{8}$
\\
$^{1}$Kavli Institute for Cosmology, University of Cambridge, Madingley Road, Cambridge, CB3 0HA, UK \\
$^{2}$Cavendish Laboratory, University of Cambridge, JJ Thomson Avenue, Cambridge, CB3 0US, UK \\
$^{3}$Steward Observatory, University of Arizona, 933 N Cherry Ave, Tucson, AZ 85721, USA \\
$^{4}$Department of Astronomy, University of California, 501 Campbell Hall \#3411, Berkeley, CA 94720, USA \\
$^{5}$Cosmic Dawn Center (DAWN) \\
$^{6}$Niels Bohr Institute, University of Copenhagen, Jagtvej 128, 2200 Copenhagen N, Denmark \\
$^{7}$Department of Physics, University of California, Santa Barbara, CA 93106, USA \\
$^{8}$Institute for Astronomy, University of Vienna, T{\"u}rkenschanzstrasse 17, A-1180 Vienna, Austria
}

\date{Accepted XXX. Received YYY; in original form ZZZ}

\pubyear{\the\year{}}

\begin{document}
\label{firstpage}
\pagerange{\pageref{firstpage}--\pageref{lastpage}}
\maketitle

\begin{abstract}
Strong \Lya{} emission observed from galaxies when the Universe is expected to be highly neutral is thought to trace large ionized regions that facilitate the transmission of \Lya{} through the IGM. In this work, we use deep \textit{JWST} \Lya{} spectroscopy to constrain the size of a candidate ionized bubble at $z\sim8.7$ in the EGS field, with a potential radius of $R_b=2$\,physical\,Mpc (pMpc) or larger. We measure a photometric galaxy density and find that the volume is a factor of $\sim2.5-3.6$ overdense, suggesting that there may be a large population of galaxies capable of creating an $R_b\sim2$\,pMpc bubble. Then, we infer the \Lya{} transmission through the IGM for galaxies in the EGS volume using our deep spectroscopy, finding $\TIGM=0.26_{-0.14}^{+0.25}$. This transmission is consistent with the average at $z\sim9$ and is mildly inconsistent with the transmission expected for an $R_b\sim2$\,pMpc bubble ($\mathcal{T}_{\textsc{igm},2\,\text{pMpc}}=0.53-0.63$), implying that such a large bubble is unlikely to be present. However, the photometric galaxy density in the EGS field is larger than in several other deep fields. This overdensity and the moderate \Lya{} transmission may be consistent with smaller, $R_b\sim0.5-1$\,pMpc bubbles in EGS. This additionally motivates the need for future wider area \Lya{} spectroscopy in EGS and other fields to obtain a more representative understanding of the sizes of ionized bubbles in the early stages of reionization, and the properties of the galaxies that create them.
\end{abstract}




\section{Introduction} \label{sec:intro}

In the first billion years after the Big Bang, the earliest galaxies emerged and reionized the Universe. During this time, these galaxies produced hydrogen-ionizing radiation that escaped their interstellar and circumgalactic media (ISM and CGM) to ionize the hydrogen in their surrounding intergalactic medium (IGM), forming local ``bubbles'' of ionized hydrogen. Over time, these ionized bubbles grew and merged until the entire IGM was fully ionized. Thus, early galaxy evolution is inextricably linked to the process of reionization, and galaxies can provide important insights into both the overall timeline and the local topology of reionization.

Over the last two decades, significant progress has been made towards observationally measuring the timeline of reionization. Measurements of the electron scattering optical depth to reionization from the cosmic microwave background (CMB) suggest that the midpoint of reionization occurred at $z = 7.7$ \citep{PlanckCollaboration2020}. While the CMB only provides an integral constraint, measurements based on observational probes sensitive to specific times in the Universe's history corroborate this picture. Studies based on quasar spectra at $z \sim 5 - 6$ have enabled constraints on the end of reionization, placing it at $z \sim 5 - 5.5$ \citep[e.g.][]{Yang2020, Zhu2021, Zhu2023, Bosman2022}, while at higher redshifts, the \Lya{} damping wing attenuation signatures observed in a few $z > 7$ quasars generally imply that the IGM is still considerably neutral \citep{Greig2017, Davies2018, Banados2018, Wang2020, Yang2020_z75_quasar}. Measurements of the neutral hydrogen fraction based on galaxies suggest the same overall timeline, where reionization is actively ongoing at $z \sim 7$ and complete at $z < 6$ (e.g. \citealt{Mason2018, Mason2025, Jung2020, Morales2021, Bolan2022, Tang2024_highz_lya, Umeda2025}; also see \citealt{Ouchi2020} for a review). However, despite significant progress in understanding when reionization occurred, charting the topology has remained elusive.

Lyman-alpha (\Lya) emission from galaxies in the heart of reionization is frequently used to measure the reionization timeline, and recently, attention has turned to leveraging \Lya{} observations to probe the reionization topology on the scales of ionized bubbles \citep{Lu2024_bubble_mapping, Nikolic2025}. As a resonant line of hydrogen, \Lya{} emitted from high-redshift galaxies is very likely to be attenuated by neutral hydrogen in the IGM, an effect that increases as the neutral hydrogen content of the IGM increases towards higher redshifts \citep[e.g.][]{Miralda-Escude1998}. However, in spite of the strong attenuation \Lya{} is expected to face in the partially neutral IGM during reionization, \Lya{} has been observed from galaxies even from galaxies at $z \gtrsim 8$ \citep{Tang2023, Bunker2023, Witten2024, Witstok2025, Witstok2025_z13}, and understanding its cause has been the topic of significant study. This reionization-era \Lya{} may be observable due to intrinsic galaxy properties such as very hard ionizing radiation fields facilitating the production of very strong \Lya{}, which is then significantly attenuated in the IGM, but still observable \citep[e.g.][]{Stark2017, Laporte2017, Endsley2021, Roberts-Borsani2023, Tang2024_highz_lya, Witstok2025, Witstok2025_z13}. Or, \Lya{} emission may escape the galaxy already significantly redshifted from systemic, as is common in bright galaxies, \citep{Erb2014, Shibuya2014, Endsley2022}. However, the visibility of \Lya{} could also be explained if \Lya{} emitters (LAEs) are embedded in large ionized regions. If LAEs inhabit large ionized bubbles, \Lya{} photons emitted at or near line center can cosmologically redshift into the damping wing while in an ionized medium before encountering neutral hydrogen, decreasing their scattering cross-section and increasing the fraction of the line that is transmitted through the IGM \citep[e.g.][]{Wyithe2005, Furlanetto2005}. In this picture, LAEs are also likely associated with overdensities of galaxies that are all collectively contributing ionizing photons towards reionizing a large volume \citep[e.g.][]{Hutter2017, Hutter2023, Dayal2018, Qin2022}, which also provides a method by which to identify candidate ionized bubbles.

Before \textit{JWST}, our knowledge of \Lya{} emission during reionization was enabled by spectroscopically following up \Lya{} break galaxy candidates identified by large ground-based telescopes \citep[e.g.][]{Stark2010, Vanzella2011, Ono2012, Pentericci2014, Mason2019_kmos, Jung2019} and narrowband imaging surveys \citep[e.g.][]{Zheng2017, Ota2017, Itoh2018, Hu2019, Goto2021, Wold2022}. These campaigns resulted in the discovery of tens of LAEs at $z \sim 6 - 8$, but only two at $z > 8$ (EGSY8p7, \citealt{Zitrin2015}; EGSY\textunderscore z910\textunderscore 44164, \citealt{Larson2022}). These $z > 8$ LAEs are extremely bright (absolute UV magnitudes of $\Muv \approx -22$) at redshifts of $z = 8.61$ and $z = 8.68$ and both are in the Extended Groth Strip (EGS) field with a physical separation of only $\sim 4$\,physical\,Mpc (pMpc). Photometric observations with \textit{HST} in the area revealed an overdensity of galaxy candidates over the entire EGS field \citep{Finkelstein2022_candels, Larson2022, Leonova2022}, and \citet{Whitler2024} identified a photometric galaxy overdensity in the vicinity of EGSY8p7 (within $\sim 5$\,arcmin) with \textit{JWST}/Near Infrared Camera \citep[NIRCam;][]{Rieke2005, Rieke2023} imaging. Taken together, the proximity of the two LAEs and the galaxy overdensity suggested that both LAEs may have inhabited the same very large ionized bubble with a minimum radius of $R_b = 2$\,pMpc at $z = 8.7$. However, recent observational constraints using a sample of $z > 6$ galaxies with \Lya{} detections (including EGSY8p7 and EGSY\textunderscore z910\textunderscore 44164 at $z = 8.7$) inferred radii of $R_b \lesssim 0.5$\,pMpc for the ionized bubble(s) associated with these objects \citep{Hayes2023}. Additionally, given the large neutral hydrogen fraction expected at $z \sim 8.7$ ($\xHI \sim 0.7 - 0.9$), a bubble with radius $R_b \gtrsim 2$\,pMpc would be unexpected by theoretical predictions \citep{Lu2024_bubble_sizes}, motivating the need to carefully quantify the extent of the candidate bubble with additional \Lya{} observations.

UV-continuum--faint galaxies are crucial for building a more complete picture of the reionization topology, but it was challenging to observe \Lya{} emission from faint galaxies before \textit{JWST}. Large ground-based facilities could detect \Lya{} in UV-continuum--bright galaxies, but bright galaxies are relatively rare, making it difficult to observe large samples to map the reionization process. Bright galaxies also tend to emit \Lya{} significantly redshifted from systemic \citep[e.g.][]{Erb2014, Shibuya2014}, likely due to large neutral hydrogen column densities through which \Lya{} must escape, making it challenging to determine if the galaxy is actually embedded in an ionized bubble or if it is surrounded by a neutral IGM that would have attenuated \Lya{} photons at line center had they been present. In contrast, faint galaxies are significantly more abundant than bright galaxies and tend to emit \Lya{} closer to systemic than bright galaxies do \citep[e.g.][]{Erb2014, Prieto-Lyon2023, Saxena2024_lya, Tang2024_z5_lya}, making them better suited for constraining the reionization topology than bright galaxies, which has been enabled for the first time by \textit{JWST}.

In this work, we characterize the transmission of \Lya{} from galaxies inside the candidate ionized bubble surrounding EGSY8p7 and EGS\textunderscore z910\textunderscore 44164 at $z = 8.7$. We obtain deep medium- and high-resolution NIRSpec observations in the rest-frame UV and rest-frame optical from JWST GO program 4287 (PIs C. Mason and D. Stark) and assemble a sample of $z = 8.7$ galaxies, then combine with measurements from the literature to constrain the \Lya{} transmission in the volume. In comparison with measurements in the field at $z \sim 9$ and expectations for the \Lya{} transmission properties of galaxies inside a very large, $R_b \gtrsim 2$\,pMpc ionized bubble that would contain both EGSY8p7 and EGS\textunderscore z910\textunderscore 4416, this enables us to examine the probability of existence for such a bubble. In combination with updated constraints on the galaxy overdensity in the volume, we can ultimately study the size of the ionized bubble at $z = 8.7$ in the EGS field for the first time.

This paper is organized as follows. In Section\ \ref{sec:observations}, we describe the spectroscopic data and analysis techniques for GO 4287 and provide a brief overview of additional \textit{HST} and \textit{JWST} data we use in this analysis. In Section\ \ref{sec:z8p7_sample}, we describe the properties of the $z = 8.7$ galaxies we have observed as part of GO 4287. In Section\ \ref{sec:overdensity}, we update the analysis of \citet{Whitler2024} to examine the photometric overdensity of galaxies in the $z = 8.7$ volume probed by EGS and its implications for the presence of a large ionized bubble, then use measurements of the transmission of \Lya{} transmission in this volume to quantify constraints on the size of the ionized bubble in Section\ \ref{sec:lya_transmission}. We then place these measurements in context with photometric overdensities in other fields in Section\ \ref{sec:discussion} and finally, summarize and conclude in Section\ \ref{sec:summary}. We adopt a flat $\Lambda$CDM cosmology with $h = 0.7$. $\Omega_m = 0.3$, and $\Omega_\Lambda = 0.7$ and all magnitudes are given in the AB system \citep{Oke1983}. We report the median and 68\,per\,cent credible interval ($16^\text{th}$ and $84^\text{th}$ percentiles) as parameter values and uncertainties.

\section{Observations and Measurements} \label{sec:observations}

\subsection{NIRSpec Observations} \label{subsec:nirspec_observations}

\begin{figure}
    \centering
    \includegraphics[width=0.48\textwidth]{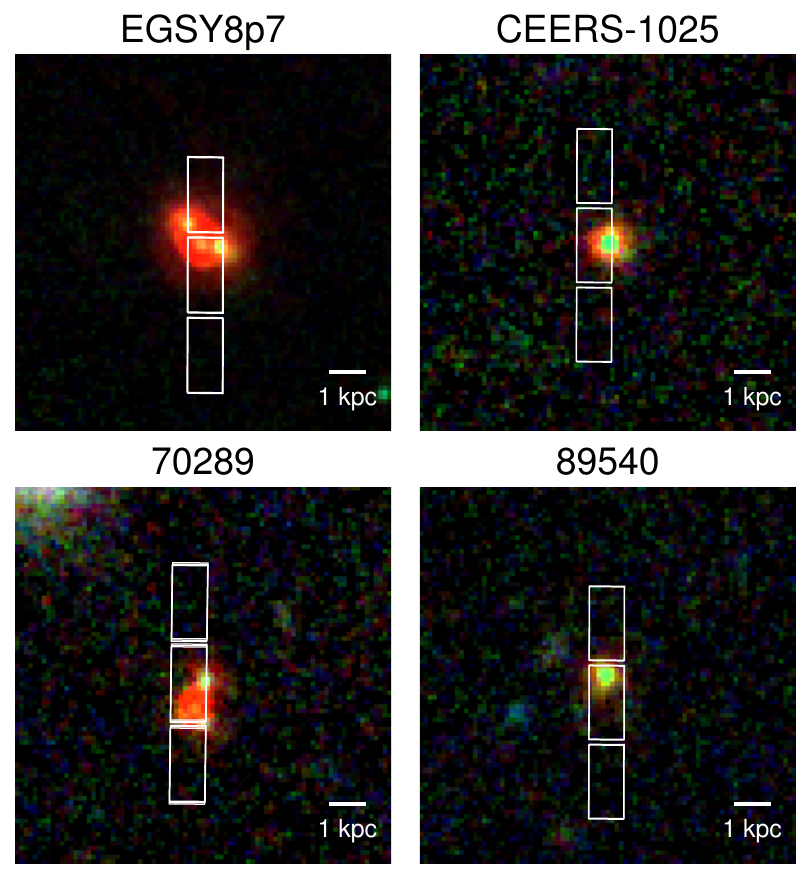}
    \caption{\label{fig:sample_shutters}F444W/F200W/F115W images of the four $z = 8.7$ objects observed by GO 4287, with the MSA slitlets from GO 4287 overlaid in white. Each image is $2.5\arcsec \times 2.5\arcsec$ and we show a 1\,kpc scale bar at the redshift of each object, measured as described in Section\ \ref{subsec:spectroscopic_measurements}.}
\end{figure}

The analysis presented in this work is based on \textit{JWST}/NIRSpec observations taken as part of JWST GO program 4287, which primarily targeted galaxies in EGS thought to trace the candidate ionized bubble at $z \sim 8.7$ and includes targets at $z \sim 7 - 8$. Spectroscopic data were taken with \textit{JWST}/NIRSpec in multi-object spectroscopy mode. We obtained data in three pointings of the NIRSpec micro-shutter array (MSA) using the disperser/filter combinations of G140H/F100LP (spectral resolution of $R \sim 2700$, targeting the wavelength of \Lya{} in our primary targets) and G395M/F290LP ($R \sim 1000$, targeting \ion{[O}{iii]}$\lambda\lambda$4959,5007 in our primary targets). For each pointing, G140H/F100LP was observed for a total exposure time of 14\,005\ s and G395M/F290LP was observed for three integrations of 16 groups for a total exposure time of 3501\ s, all using the NRSIRS2 readout mode. Each MSA slitlet was composed of three shutters and observations were taken using a three-point nod pattern with one exposure at each nod position.

After the assignment of the observation position angles (PAs), we used the NIRSpec MSA Planning Tool to optimize the MSA pointing centers and shutter configurations in order to maximize the number of primary targets that potentially inhabited ionized bubbles at $z \gtrsim 7$. We identified 88 high-priority targets using the photometrically selected samples of \citet{Chen2024} and \citet{Whitler2024}, who searched for overdensities of galaxies (and therefore candidate ionized bubbles) at $z = 7.2, 7.5,\ \text{and}\ 8.7$. We then required each object to be observed at the possible wavelengths of both \ion{Ly}{$\alpha$} and \ion{[O}{iii]}$\lambda\lambda$4959,5007 based on their photometric redshift probability distributions. After applying these requirements, we assigned slits to 32 of our highest priority targets, including seven targets selected by \citet{Whitler2024} to lie at redshifts of $z \sim 8.4 - 9.1$, potentially tracing an ionized bubble at $z = 8.7$. We then filled the remaining space on the MSA with 127 Lyman break galaxy candidates primarily selected to be at $z \geq 5$ (F814W dropouts) but down to $z \geq 3$ as necessary to fill the MSA, resulting in a total of 159 objects assigned to shutters. We then reduce the 2D spectra following the methods described by \citet{topping2024_rcxj}, which uses the standard \texttt{jwst} reduction pipeline\footnote{\url{https://github.com/spacetelescope/jwst}} \citep{Bushouse2025} in combination with custom algorithms. To obtain the 1D spectra, we fit the spatial profile of each 2D spectrum with a Gaussian, then perform a boxcar extraction centered on the mean of the Gaussian with a typical width of five pixels (0.5\,arcsec) in the spatial direction.

In this work, we primarily present four galaxies at $z = 8.7$ observed as part of GO 4287 (IDs EGSY8p7, CEERS-1025, 70289, and 89540). We show the MSA slitlets of these objects overlaid on F444W/F200W/F115W RGB images in Figure\ \ref{fig:sample_shutters}. All of these objects have been previously observed with the NIRSpec prism ($R \sim 100$), G140M, and G395M ($R \sim 1000$) gratings by the Cosmic Evolution Early Release Science (CEERS) program \citep[PI S. Finkelstein;][]{Finkelstein2025}; with the prism and G395M grating by the Red Unknowns: Bright Infrared Extragalactic Survey (RUBIES) program \citep[PIs A. de Graaff and G. Brammer;][]{deGraaff2024}; and/or the prism with as part of the CANDELS-Area Prism Epoch of Reionization Survey (CAPERS, PI M. Dickinson). To maximize signal-to-noise, we reduce the G395M spectra from these programs using the same methods we use to reduce the GO 4287 observations, then stack the GO 4287 spectra with the spectra from CEERS and/or RUBIES for EGSY8p7, CEERS-1025, and 70289 weighted by exposure times. ID 89540 was also observed by RUBIES in G395M, but its RUBIES spectrum has evidence of an artifact elevating the background noise, so we use only the GO 4287 spectrum for this object. The EGS field also has a large amount of prism spectroscopy observed as part of CEERS. While we do not include these observations in our \Lya{} measurements due to the challenges of using low-resolution prism measurements for \Lya{} \citep[e.g.][]{Keating2024, Chen2024}, we also reduce and analyze these datasets in order to identify galaxies that have been spectroscopically confirmed at our redshifts of interest ($z \sim 8.6 - 8.8$).

\subsection{NIRCam Observations} \label{subsec:nircam_parallels}

\begin{table*}
    \centering
    \caption{The line properties of the four objects at $z = 8.7$ that have been observed by GO 4287. We report coordinates, systemic redshifts, \ion{[O}{iii]$\lambda\lambda$4959,5007} and \ion{H}{$\beta$} rest-frame EWs, rest-UV line detections (besides \Lya), and rest-optical line detections (besides \ion{[O}{iii]} and \ion{H}{$\beta$}) for all objects. We also report \Lya{} redshifts, velocity offsets, fluxes, and rest-frame EWs for the two objects with \Lya{} detections, and report $3\sigma$ upper limits on \Lya{} fluxes and rest-frame EWs for the other two.}
    \label{tab:line_properties}
    \renewcommand{\arraystretch}{1.4}
    \begin{tabular}{lcccc}
    \hline
    & EGSY8p7 & CEERS-1025 & 70289 & 89540 \\ \hline\hline
    RA [deg] & 215.03539 & 214.96753 & 214.84477 & 214.96869 \\
    Dec [deg] & +52.89067 & +52.93296 & +52.89211 & +52.92965 \\
    $z_\text{sys}$ & $8.677 \pm 0.0001$ & $8.716 \pm 0.0003$ & $8.687 \pm 0.0002$ & $8.716 \pm 0.0006$ \\
    $z_{\Lya}$ & $8.683_{-0.001}^{+0.002}$ & $8.724_{-0.001}^{+0.001}$ & --- & --- \\
    $\Delta v_{\Lya}$ [km\,s$^{-1}$] & $164_{-33}^{+66}$ & $251_{-33}^{+33}$ & --- & --- \\
    $f_{\Lya}$ [10$^{-18}$\,erg\,s$^{-1}$\,cm$^{-2}$] & $1.67_{-0.42}^{+0.40}$ & $0.54_{-0.29}^{+0.31}$ & $< 1.52$ & $< 1.38$ \\
    $\text{EW}_{0, \Lya}$ [\AA] & $7.6 \pm 2.2$ & $3.5 \pm 2.2$ & $< 16.9$ & $< 28.9$ \\
    $\text{EW}_{0, \text{H}\beta}$ [\AA] & $257_{-16}^{+15}$ & $178_{-33}^{+35}$ & $88_{-13}^{+13}$ & $67_{-37}^{+33}$ \\
    $\text{EW}_{0, \ion{[O}{iii]}}$ [\AA] & $2330_{-78}^{+65}$ & $1065_{-53}^{+52}$ & $587_{-28}^{+31}$ & $523_{-80}^{+52}$ \\
    UV lines & \ion{N}{iv]}$\lambda$1486, \ion{C}{iv}$\lambda$1550 & \ion{N}{v}$\lambda$1243 & --- & --- \\
    \multirow{3}{*}{Optical lines} & \ion{[O}{ii]}$\lambda$3727, \ion{[Ne}{iii]}$\lambda$3869, \ion{H}{$\zeta$}+ & \ion{[O}{ii]}$\lambda$3727, \ion{[Ne}{iii]}$\lambda$3869, \ion{H}{$\zeta$}+ & \ion{[O}{ii]}$\lambda$3727, \ion{[Ne}{iii]}$\lambda$3869, & \multirow{3}{*}{---} \\
    & \ion{He}{i}$\lambda$3889, \ion{H}{$\epsilon$}+\ion{[Ne}{iii]}$\lambda$3967, & \ion{He}{i}$\lambda$3889, \ion{H}{$\epsilon$}+\ion{[Ne}{iii]}$\lambda$3967,3 & \ion{H}{$\epsilon$}+\ion{[Ne}{iii]}$\lambda$3967 & \\
    & \ion{H}{$\delta$}, \ion{H}{$\gamma$}, \ion{[O}{iii]}$\lambda$4363 & \ion{H}{$\delta$}, \ion{H}{$\gamma$}, \ion{[O}{iii]}$\lambda$4363 & \ion{H}{$\gamma$}, \ion{[O}{iii]}$\lambda$4363 & \\ \hline
    \end{tabular}
\end{table*}

For each of the three NIRSpec pointings, we also obtained coordinated NIRCam imaging parallels designed to complement and enhance existing imaging in the EGS field. For two of our three NIRSpec pointings, the NIRCam parallels did not overlap with any existing imaging. For these two pointings (a total imaging area of $\sim 18$\ arcmin$^2$), we observed in three pairs of short/long wavelength NIRCam filters: F115W/F444W, F150W/F356W, and F200W/F277W. For our third pointing, the NIRCam parallel partially overlapped with existing NIRCam imaging from CEERS in F115W, F150W, F200W, F277W, F356W, F410M, and F444W. For this pointing, we obtained deep F090W and F480M imaging (exposure time of 15\,880\ s) to expand the wavelength coverage in the $\sim 3$\ arcmin$^2$ area of overlap. In particular, the F480M filter contains the strongest rest-optical emission lines accessible by NIRCam at $z \sim 8.7$, \ion{[O}{iii]} and \ion{H}{$\beta$}. This enables more precise photometric redshift measurements and characterization of the rest-frame optical spectral energy distributions (SEDs) of galaxies at $z = 8.7$.

In addition to the CEERS NIRCam imaging in F115W, F150W, F200W, F277W, F356W, F410M, F444W, and our own parallel imaging, we also analyze F090W imaging from JWST GO program 2234 (PI E. Ba\~nados). This F090W imaging covers the same area as CEERS and provides an additional dropout filter for our photometric selection of galaxy candidates at $z \sim 8.7$ (Section\ \ref{sec:overdensity}). We reduce all of these imaging data and combine all exposures for a given filter into one mosaic per filter, then perform source detection and photometry on these final mosaics. For a description of our imaging reduction, detection methods, and photometric measurements, we refer the reader to \citet{Endsley2023, Endsley2024} \citep[also see][]{Whitler2024}.

\subsection{Spectroscopic Measurements} \label{subsec:spectroscopic_measurements}

We now present the spectroscopic redshift and line measurements we perform on the objects observed as part of GO 4287. We first use strong rest-frame optical emission lines to measure systemic spectroscopic redshifts for $z \geq 4$ objects that have multiple significant line detections (typically \ion{H}{$\alpha$}, \ion{H}{$\beta$}, and/or \ion{[O}{iii]}$\lambda\lambda$4959,5007). We then focus on the subset of galaxies at that are observed by GO 4287 at $z = 8.7 \pm 0.1$. As we are primarily interested in using the \Lya{} emission observed from these objects to probe a potential ionized bubble in this volume, we focus on measuring or placing upper limits on the \Lya{} line fluxes and equivalent widths (EWs), along with \Lya{} velocity offsets for the objects that have \Lya{} detections. We also measure and briefly discuss the \ion{[O}{iii]} and \ion{H}{$\beta$} fluxes and equivalent widths of each $z = 8.7$ object in Section\ \ref{sec:z8p7_sample}.

To measure spectroscopic redshifts, we begin by visually inspecting all of our 2D G395M spectra for rest-optical emission lines and identify 67 objects with multiple rest-optical emission line detections and an additional 14 objects with only one line detection. We do not search for \Lya{} breaks, as we do not expect the rest-UV continuum to be significantly detected in our G140H data. After visually identifying strong lines, we then measure spectroscopic redshifts by individually fitting Gaussian profiles to each of the strong lines available in the spectra (\ion{H}{$\alpha$}, \ion{H}{$\beta$}, \ion{[O}{iii]}$\lambda$4959, and/or \ion{[O}{iii]}$\lambda$5007), then using the mean of the fitted Gaussian to calculate the redshift. We perform this procedure 1000 times for each line after resampling the spectra 1000 times assuming Gaussian flux errors, then weight each redshift measurement by the signal-to-noise (S/N) of the integrated emission line flux and take the weighted average and standard deviation as the final spectroscopic redshift and uncertainty. We find 30 objects at redshifts of $z \geq 6$, 19 at redshifts of $4 \leq z < 6$, and 18 at $z < 4$. Of the 30 objects at $z \geq 6$, seven lie between $z = 8 - 9$ and four are at $z = 8.6 - 8.8$ with IDs EGSY8p7 \citep{Zitrin2015}, CEERS-1025, 70289, and 89540. We will focus on these four $z \sim 8.7$ galaxies for the remainder of this work: EGSY8p7, CEERS-1025, 70289, 89540.

After measuring systemic spectroscopic redshifts, $z_\text{sys}$, we search for \Lya{} emission lines in the four objects at $z \sim 8.7$. We visually identify \Lya{} in two objects (CEERS-1025 for the first time, and EGSY8p7), and place $3\sigma$ upper limits on the remaining two. All fluxes and EWs are reported in Table\ \ref{tab:line_properties}. For objects with \Lya, we numerically estimate uncertainties for line fluxes and EWs. We generate 1000 realizations of the G140H spectrum by perturbing the fluxes in each wavelength bin by the errors, assuming they are Gaussian and uncorrelated with neighboring pixels. We then measure the continuum around \Lya{} by stacking every wavelength pixel of the spectrum between rest-frame $1250 - 1300$\,\AA{} (ensuring that these wavelengths do not overlap the detector gap) for each of the 1000 realizations, which results in $\text{S/N} \gtrsim 2.5$ continuum measurements for each object.

To measure \Lya{} line fluxes, we subtract the stacked continuum measurement from the full spectrum, then directly integrate the continuum-subtracted spectrum, using an integration range from $\lambda = [1215.67 \times (1 + z_\text{sys})]$\,\AA{} to the first wavelength redward of $[1215.67 \times (1 + z_\text{sys})]$\,\AA{} where the observed spectrum becomes negative (resulting in $\Delta v \sim 400 - 600$\,km\,s$^{-1}$). We then take the median, 16$^\text{th}$, and 84$^\text{th}$ percentiles of the integrated line fluxes of the 1000 resampled spectra as the \Lya{} line flux and uncertainties. We adopt a similar method of direct integration to measure EWs, but now report the mean and standard deviation of the resulting distribution of \Lya{} EWs as the value and uncertainty, as symmetric errors are required for our later inference of the transmission of \Lya{} through the IGM (Section\ \ref{sec:lya_transmission}). Finally, we measure \Lya{} velocity offsets using the offset between the expected systemic wavelength of \Lya{}, and the wavelength of the spectral pixel with the peak flux of the line.

For objects without \Lya{} detections, we directly integrate the 1D error spectrum between $\Delta v = -1000$ and $+1000$\,km\,s$^{-1}$ relative to the systemic redshift, where the window is chosen to ensure that we capture the entire \Lya{} line profile expected at $z \gtrsim 6$ \citep[e.g.][]{Saxena2024_lya, Tang2024_z5_lya, Lin2024}. We then combine the integrated error spectrum with the stacked continuum measurement to place $3\sigma$ upper limits on the fluxes and EWs.

To measure the fluxes of \ion{H}{$\beta$} and both components of the \ion{[O}{iii]}$\lambda\lambda$4959,5007 doublet, we fit and integrate Gaussian functions for each line for 1000 realizations of the spectrum, with line centers set by the systemic spectroscopic redshift of the object under consideration. We note that while two of the objects in our sample (EGSY8p7 and 70289) have $\text{S/N} \gtrsim 3.5$ detections of their rest-optical continuum in our G395M spectra after stacking the wavelength pixels in the vicinity of \ion{[O}{iii]} and \ion{H}{$\beta$}, the remaining two objects are not detected. Thus, rather than using continuum fluxes measured directly from the spectra to measure EWs, we rescale the G395M spectrum of each object to match the normalization of the object's NIRCam photometry, then combine the rescaled spectra with continuum measurements inferred from SED models of the photometry. In detail, we renormalize the spectra by calculating synthetic photometry from the G395M spectra in relevant NIRCam filters (F356W, F410M, F444W), taking the ratios in each filter of the observed NIRCam fluxes and the synthetic fluxes derived from the spectra, then rescaling the entire G395M spectrum by the average of the ratios.

\begin{figure*}
    \centering
    \includegraphics[width=0.48\textwidth]{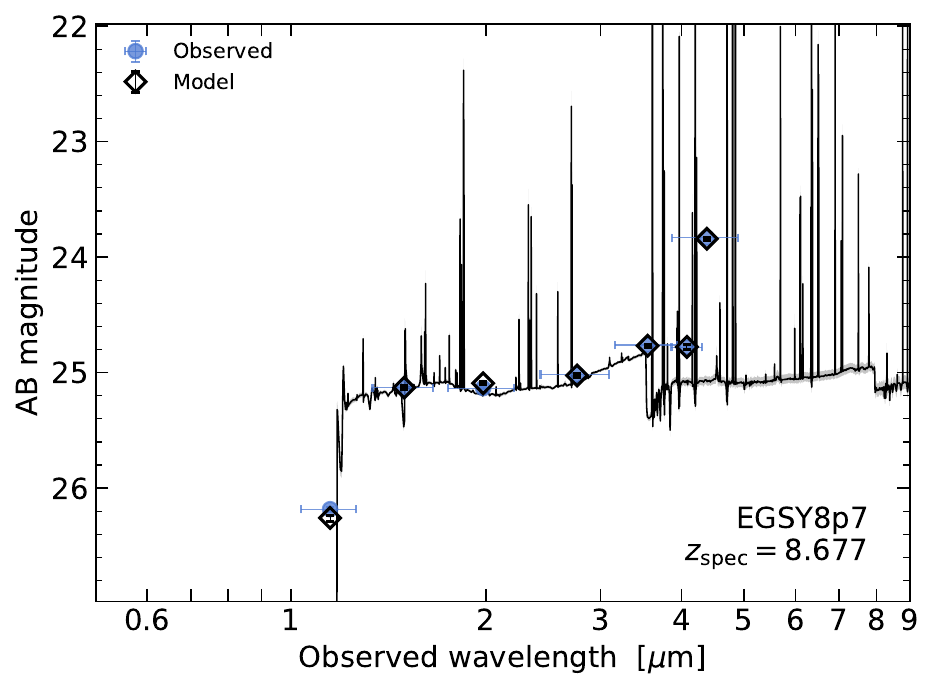}
    \includegraphics[width=0.48\textwidth]{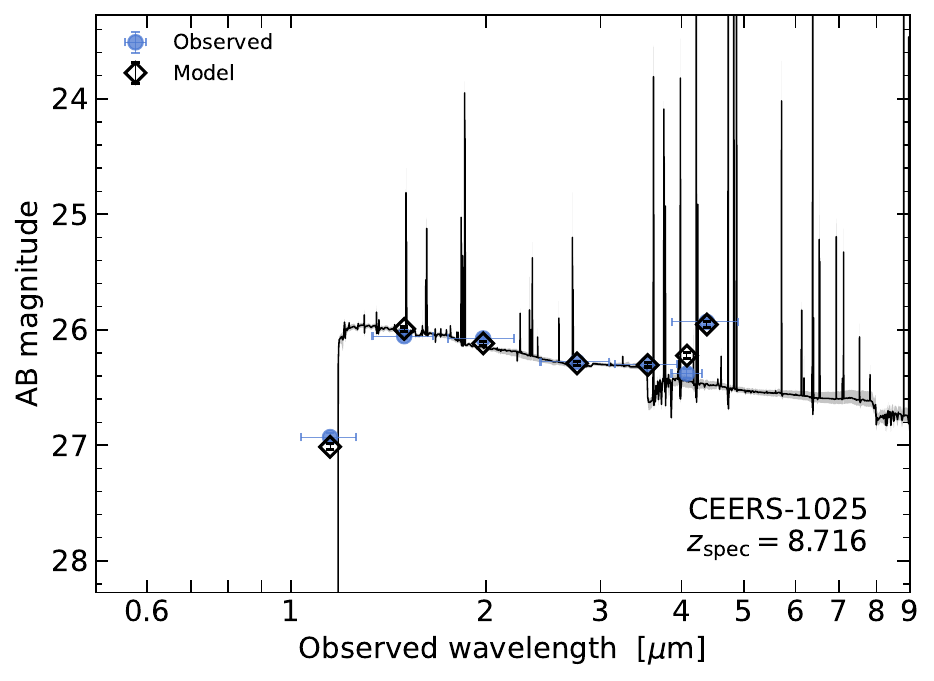}
    \includegraphics[width=0.48\textwidth]{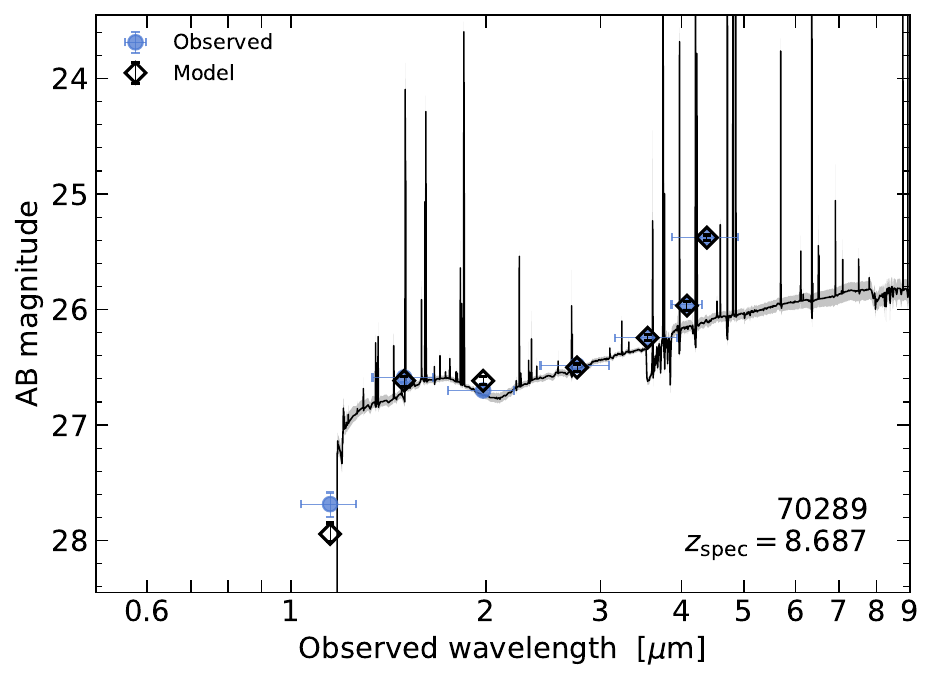}
    \includegraphics[width=0.48\textwidth]{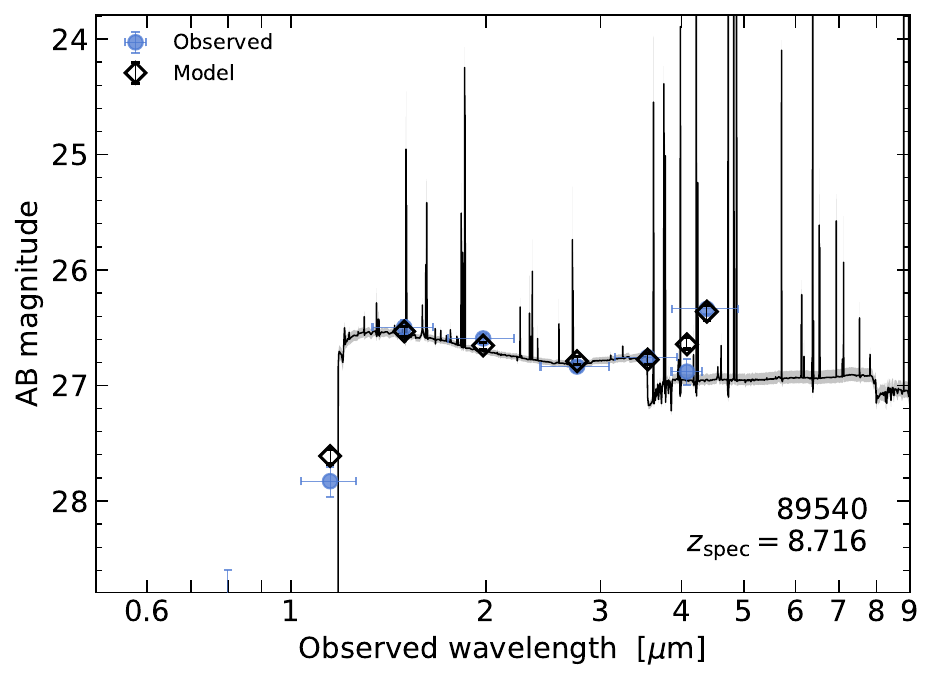}
    \caption{\label{fig:nircam_seds}The NIRCam SEDs of the four $z = 8.7$ objects observed by GO 4287. We show the observed photometry as blue circles, the model photometry as open black diamonds, and the model spectra with uncertainies as black lines and grey shaded regions. All four galaxies are moderately to extremely bright, with absolute UV magnitudes ranging from $-22.1 \lesssim \Muv \lesssim -20.7$ and are inferred to have moderate to high ionizing photon production efficiencies of $\xi_\text{ion} = 10^{25.3 - 25.8}$\,Hz\,erg$^{-1}$.}
\end{figure*}

\section{Sample Properties} \label{sec:z8p7_sample}

\begin{figure*}
    \includegraphics[width=\textwidth]{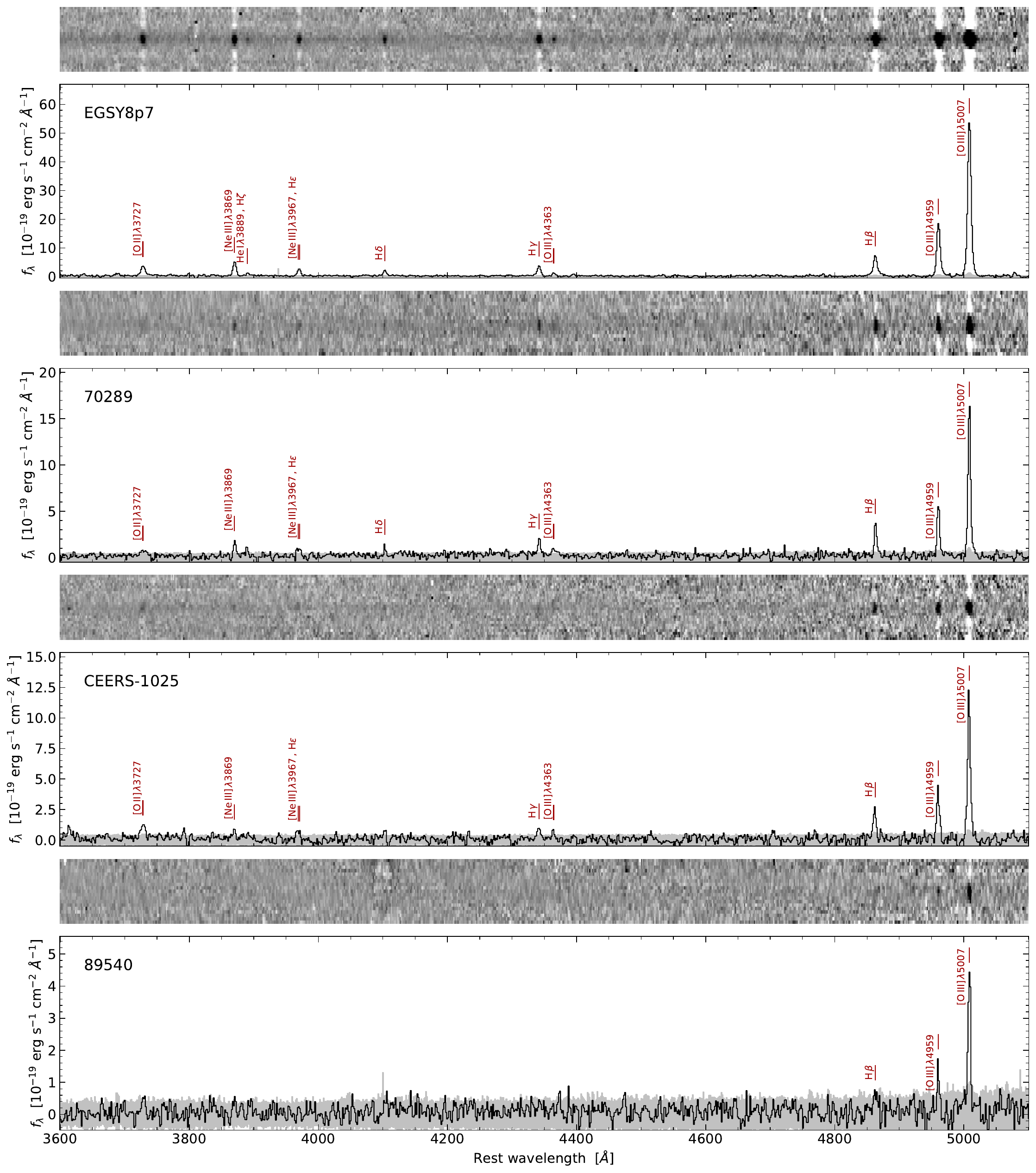}
    \caption{\label{fig:g395m_spectra}The G395M spectra, probing the rest-frame optical, of the four objects at $z = 8.7$ that were observed by GO 4287. All show strong \ion{[O}{iii]} emission, and several have multiple additional Balmer line detections. Of these four objects, two are observed to have \Lya{} emission (EGSY8p7 and CEERS-1025, see Figures\ \ref{fig:69787_uv_lines} and \ref{fig:87873_uv_lines}).}
\end{figure*}

\begin{figure}
    \centering
    \includegraphics[width=\columnwidth]{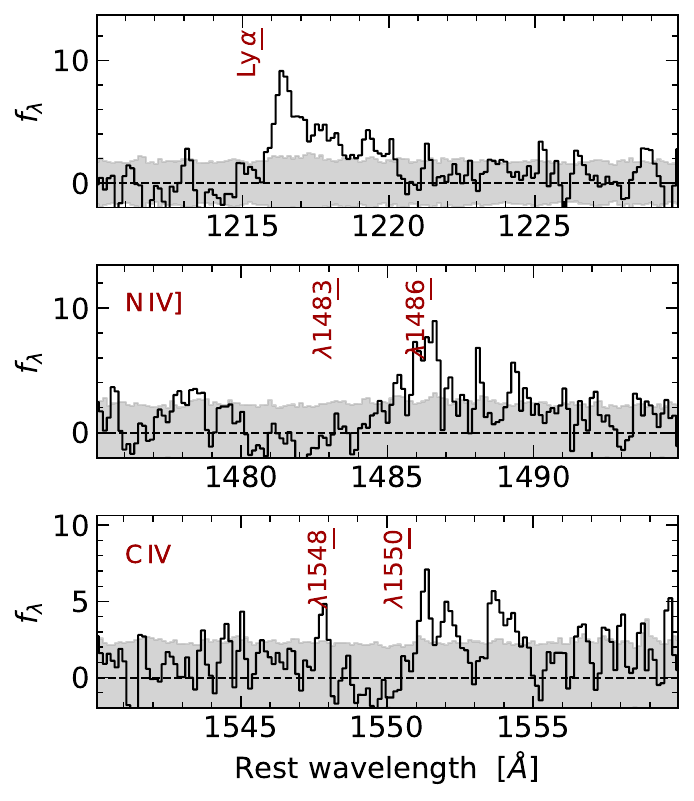}
    \caption{\label{fig:69787_uv_lines}The rest-frame UV lines detected in the G140H spectrum of EGSY8p7 from GO 4287. We detect \Lya{} and the red components of the \ion{N}{iv]}$\lambda\lambda$1483,1486 and \ion{C}{iv}$\lambda\lambda$1548,1550 doublets. We measure an equivalent width for \Lya{} of $\text{EW}_{0, \Lya} \sim 8$\,\AA{} and a velocity offset of $\Delta v = 164$\,km\,s$^{-1}$.}
\end{figure}

The EGS field has long been studied as a potentially unusual site of intense early star formation activity at $z = 8.7$, hosting two bright LAEs \citep{Zitrin2015, Larson2022} along with evidence of a galaxy overdensity and a candidate ionized bubble at $z = 8.7$ \citep{Finkelstein2022_candels, Larson2022, Tang2023, Whitler2024}. In the last few years, JWST imaging and spectroscopy has revealed an abundance of moderately bright galaxies ($\Muv \lesssim -20$) in EGS at $z = 8.7$, several of which have detections of high-ionization emission lines \citep[e.g.][]{Larson2023, Topping2025_CIV_N, Tang2025}, which may also indicate the presence of an active galactic nucleus \citep[AGN,][]{Larson2023}. If these galaxies are producing copious amounts of ionizing photons either by star formation and/or AGN activity, they may then be capable of creating a very large ionized bubble as early as $z \sim 9$, and in this section, we examine the star-forming and ionizing properties of the four $z = 8.7$ galaxies that have been observed as part of GO 4287 in the context of the formation of an early ionized bubble and the visibility (or lack thereof) of their \Lya{} emission.

To infer the physical properties of these galaxies, we model their NIRCam SEDs with the BayEsian Analysis of GaLaxy sEds \citep[\beagle;][]{Chevallard2016} code. \beagle{} is underpinned by an updated version of the \citet{Bruzual2003} stellar population synthesis models \citep{Vidal-Garcia2017} and the nebular emission models (both line and continuum) of \citet{Gutkin2016}, which were in turn modelled with the photoionization code \texttt{Cloudy} \citep{Ferland2013}. We adopt a \citet{Chabrier2003} stellar initial mass function with a mass range of $0.1 - 300$\,\Msun and a Small Magellanic Cloud (SMC) dust law \citep{Pei1992}, and model the attenuation by the IGM using the prescription of \citet{Inoue2014}. For the star formation history, we adopt a two-component parametrization consisting of a delayed exponential (i.e. delayed-$\tau$) model at early times plus a recent episode of constant star formation,\footnote{The functional form of this star formation history is \[ \text{SFR}(t) \propto \begin{cases} 1, & t < t_\text{const} \\ t\text{e}^{-t / \tau}, & t_\text{const} <  t < t_\text{start}\end{cases} \]} as was first introduced by \citet{Endsley2024}. We place a log-uniform prior on the maximum stellar age (i.e. the start time of the star formation history, $t_\text{start}$) with a range of $20$\,Myr to the age of the Universe at the systemic redshift of the source under consideration, a uniform prior on the \textit{e}-folding time of the delayed-$\tau$ component ($\tau$) between $1\,\text{Myr} - 30\,\text{Gyr}$, a log-uniform prior on the time the delayed-$\tau$ model ends and the constant component starts ($t_\text{const}$) from $1 - 20$\,Myr, and a log-uniform prior on the specific star formation rate of the constant component ranging from $10^{-5} - 10^{3}$\,Gyr$^{-1}$. We also model the stellar mass ($M_* = 10^5 - 10^{12}$\,\Msun), stellar metallicity ($Z_* = 0.006 - 0.5$\,Z$_\odot$, where Z$_\odot = 0.01524$; \citealt{Caffau2011}), ionization parameter ($U = 0.0001 - 0.1$), and the $V$-band optical depth due to dust attenuation ($\tau_\textsc{v} = 0.001 - 5$), all with log-uniform priors. Finally, we assume that the total interstellar gas- and dust-phase metallicity is the same as the stellar metallicity with dust-to-gas mass ratios allowed to vary with a uniform prior between $\xi_d = 0.1 - 0.5$ (noting that \beagle{} self-consistently models the effects of dust depletion). We show the resulting NIRCam SEDs and model results in Figure\ \ref{fig:nircam_seds} and use the inferred physical properties from these models for the remainder of this work, after confirming that the values of $\xi_\text{ion}$ we infer from the SED models are consistent with the values implied by the \ion{H}{$\beta$} fluxes we directly observe in our spectra (using the dust optical depths from the SED models and the same SMC dust extinction curve to apply a dust correction). We provide a brief overview of the spectroscopic and SED model-inferred physical properties of the four $z = 8.7$ galaxies in the following sections, but refer to Table\ \ref{tab:line_properties} and to Table\ \ref{tab:phot_sample_obs_props} for a summary of SED model-based properties not discussed here.

\subsection{EGSY8p7} \label{subsec:egsy8p7}

EGSY8p7 \citep[also CEERS-1019, and ID 69787 in][]{Whitler2024} is an extremely bright ($m_\text{F150W} = 25.1$) galaxy at $z = 8.677$, which was first photometrically identified by \citet{Roberts-Borsani2016} as a bright $z \sim 7 - 9$ candidate with a photometric excess in the \textit{Spitzer}/Infrared Array Camera 4.5\,$\mu$m channel that implied very strong \ion{[O}{iii]}+\ion{H}{$\beta$} emission. EGSY8p7 was later spectroscopically confirmed via Keck/MOSFIRE observations of \Lya{} (and was the first LAE known at $z > 8$) at a redshift of $z = 8.683$ by \citet{Zitrin2015}. This object has also been studied extensively with both imaging and medium-resolution spectroscopy from \textit{JWST} as part of CEERS, RUBIES, and CAPERS, and in this work, we present the first high-resolution ($R \sim 2700$) spectrum in the rest-frame UV taken as part of GO 4287.

In the rest-UV, EGSY8p7 shows \Lya{} emission with $\EW_{0, \Lya} = 7.6 \pm 2.2$\,\AA). With our high resolution rest-UV spectrum, we find that \Lya{} is offset redward of the systemic redshift by $\Delta v_{\Lya} = 164_{-33}^{+66}$\,km\,s$^{-1}$, a slightly smaller \Lya{} velocity offset than was found by \citet{Tang2024_highz_lya}. We attribute this difference to the more precise constraints enabled by the higher resolution ($R \sim 2700$) G140H spectrum compared to the previously available $R \sim 1000$ G140M spectrum. As shown in Figure\ \ref{fig:69787_uv_lines}, EGSY8p7 also shows the red components of two high-ionization rest-UV lines, \ion{N}{iv]}$\lambda\lambda$1483,1486 \citep[previously discussed by][]{Larson2023, Isobe2023,Topping2025_CIV_N} and \ion{C}{iv}$\lambda\lambda$1548,1550 \citep{Topping2025_CIV_N}, requiring photon energies of $\gtrsim 47$\,eV and potentially indicate of a very hard ionizing spectrum.

In the rest-optical, EGSY8p7 has a very large suite of line detections (see Figure\ \ref{fig:g395m_spectra}): hydrogen Balmer lines from \ion{H}{$\beta$} through \ion{H}{$\zeta$}, both nebular ($\lambda\lambda$4959,5007) and auroral ($\lambda$4363) \ion{[O}{iii]}, nebular \ion{[O}{ii]}$\lambda$3727 and \ion{[Ne}{iii]}$\lambda$3869, which have been previously discussed by \citet{Tang2023, Larson2023}. Fitting the strongest lines (\ion{H}{$\beta$}, \ion{[O}{iii]}$\lambda$4959, and \ion{[O}{iii]}$\lambda$5007) with Gaussian profiles yields a systemic spectroscopic redshift of $z_\text{sys} = 8.677$. Given its observed F150W magnitude, this corresponds to an absolute UV magnitude of $\Muv = -22.1$, the brightest in our sample. We measure very high \ion{H}{$\beta$} and nebular \ion{[O}{iii]} EWs of $\EW_{0, \ion{H}{$\beta$}} = 257_{-16}^{+15}$\,\AA{} and $\EW_{0, \ion{[O}{iii]}} = 2330_{-78}^{+65}$\,\AA, which are among the most extreme \ion{[O}{iii]} and \ion{H}{$\beta$} EWs that have been directly observed at $z \sim 6 - 9$ \citep{Matthee2023, Meyer2024, Roberts-Borsani2024} and above the median of the photometrically inferred \ion{[O}{iii]}+\ion{H}{$\beta$} EW distribution at $z \sim 7 - 9$ \citep{Endsley2024, Begley2025}. Such strong nebular line emission is linked to hard ionizing radiation fields that may be able to contribute significantly to ionizing the nearby IGM, and is empirically correlated with stronger \Lya{} than in galaxies with weaker \ion{[O}{iii]}+\ion{H}{$\beta$} emission \citep[e.g.][]{Chen2024, Tang2024_z5_lya}, which may be helping to facilitate the detection of $\text{EW}_{0, \Lya} \sim 7.5$\,\AA{} \Lya{} emission in this object.

From our SED models of the NIRCam photometry of EGSY8p7, we infer properties consistent with a recent, intense burst of star formation that is producing copious amounts of ionizing photons, but possibly for only a short amount of a time. We infer a large ionizing photon production efficiency of $\xi_\text{ion} = 10^{25.8}$\,Hz\,erg$^{-1}$, consistent with the picture implied by the strong nebular line emission and high-ionization UV lines we observe in the spectrum, where EGSY8p7 has an intense ionizing radiation field. This suggests that EGSY8p7 may have very strong intrinsic \Lya{}, facilitating its observability. Additionally, EGSY8p7 may be able to contribute significant quantities of ionizing photons to creating a large ionized region, as well as produce very strong \Lya{}. However, we also infer a young mass-weighted age of $\sim 13$\,Myr. This suggests that though EGSY8p7 is producing ionizing photons extremely efficiently, the most recent episode of star formation is short enough such that the observable stellar population may not have had sufficient time to contribute significantly to reionizing a large bubble. However, this does not necessarily preclude the creation of an ionized bubble by a series of preceding star formation episodes that are similarly efficient at producing ionizing photons.

\subsection{CEERS-1025} \label{subsec:87873}

CEERS-1025 is a bright, $m_\text{F150W} = 26.1$ galaxy at a redshift of $z = 8.717$, which was first observed and spectroscopically confirmed with medium resolution NIRSpec observations \citep{Nakajima2023, Tang2023} from the CEERS program. CEERS-1025 lies 1.4\,pMpc away from EGSY8p7, well within the confines of a very large, $R_b = 2$\,pMpc ionized region. Though fainter than EGSY8p7, this galaxy shows a similarly large suite of rest-optical lines, including \ion{H}{$\beta$}, \ion{H}{$\gamma$}, nebular and auroral \ion{[O}{iii]}, \ion{[O}{ii]}, and \ion{[Ne}{iii]}. After fitting \ion{[O}{iii]}$\lambda\lambda$4959,5007 and \ion{H}{$\beta$}, we find a systemic spectroscopic redshift of $z_\text{sys} = 8.716$. This corresponds to an absolute UV magnitude of $\Muv = -21.2$, making CEERS-1025 the second brightest object in both our spectroscopic and photometric (see Section\ \ref{sec:overdensity}) samples. We measure smaller EWs for \ion{H}{$\beta$} and \ion{[O}{iii]} in CEERS-1025 than we do in EGSY8p7 ($\EW_{0, \ion{H}{$\beta$}} = 178_{-33}^{+35}$\,\AA{} and $\EW_{0, \ion{[O}{iii]}} = 1065_{-53}^{+52}$\,\AA), but these EWs still fall at the high end of expected \ion{[O}{iii]}+\ion{H}{$\beta$} from both spectroscopic and photometric observations.

\begin{figure}
    \centering
    \includegraphics[width=\columnwidth]{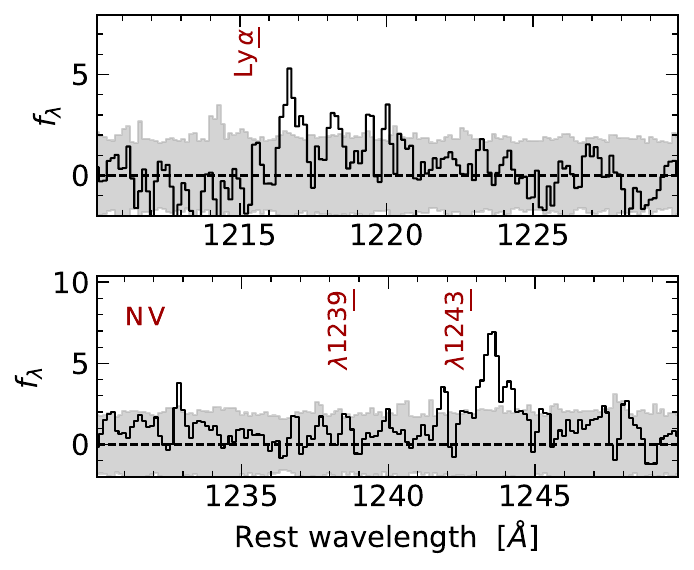}
    \caption{\label{fig:87873_uv_lines}The rest-frame UV lines detected in the G140H spectrum of CEERS-1025 from GO 4287. We detect \Lya{} and the blue component of the high-ionization \ion{N}{v}$\lambda\lambda$1239,1243 doublet. We measure an equivalent width for \Lya{} of $\text{EW}_{0, \Lya} = 3$\,\AA{} redshifted from systemic by $\Delta v = 251$\,km\,s$^{-1}$. \citet{Tang2025} provides a detailed discussion of the implications of the \ion{N}{v} detection for this object.}
\end{figure}

Our rest-UV observations from GO 4287 also cover the wavelength of \Lya{} for the first time, and we detect \Lya{} along with the red component of the high-ionization \ion{N}{v}$\lambda\lambda$1239,1243 doublet (Figure\ \ref{fig:87873_uv_lines}). For \Lya, we measure a rest-frame EW of $\text{EW}_{0, \Lya} = 3.5 \pm 2.2$\,\AA{} and a velocity offset of $\Delta v_{\Lya} = 251_{-33}^{+33}$. We refer to \citet{Tang2025} for a detailed discussion of the \ion{N}{v} line and other high-ionization UV lines; however, we note that, similarly to EGSY8p7, the presence of \ion{N}{v} suggests that CEERS-1025 has a hard radiation field with the presence of photons with energies $\gtrsim 77$\,eV. As for EGSY8p7, we infer a mass-weighted age of $\sim 13$\,Myr. However, we infer a markedly lower -- though still high -- ionizing photon production efficiency of $\xi_\text{ion} = 10^{25.3}$\,Hz\,erg$^{-1}$ in comparison with EGSY8p7. This suggests that this source may not contribute as much ionizing flux towards creating an ionized bubble over the course of its lifetime.

\subsection{Galaxies without detected \texorpdfstring{\Lya}{Lya} emission: 70289 and 89540} \label{subsec:70289_89540}

ID 70289 is a bright galaxy (apparent magnitude of $m_\text{F150W} = 26.6$), which has also been previously observed as part of RUBIES with the ID RUBIES-980841. The SED of 70289 increases towards rest-optical wavelengths (see bottom left panel of Figure\ \ref{fig:nircam_seds}), so that in the G395M data from GO 4287 that probes the rest-optical, we detect \ion{H}{$\beta$}, \ion{H}{$\gamma$}, \ion{H}{$\delta$}, \ion{[O}{iii]}$\lambda\lambda$4959,5007, \ion{[O}{iii]}$\lambda$4363, \ion{[O}{ii]}$\lambda$3727, \ion{[Ne}{iii]}$\lambda$3967 blended with \ion{H}{$\epsilon$}, \ion{[Ne}{iii]}$\lambda$3869, and tentative \ion{[O}{ii]}$\lambda$3727. From these lines, we measure a redshift of $z_\text{spec} = 8.687$, which implies an absolute magnitude of $\Muv = -20.7$ (given the observed F150W magnitude of 70289) and placing this object $\sim 1.9$\,pMpc away from EGSY8p7. We do not detect any rest-UV lines in our G140H observations, but place a $3\sigma$ upper limit on the EW of \Lya{} of $\text{EW}_{0, \Lya} < 17.0$\,\AA. However, we highlight that this object is fainter in the rest-UV continuum than either EGSY8p7 or CEERS-1025, and we cannot rule out \Lya{} of similarly low-EW as the two objects with \Lya{} detections. In marked contrast to EGSY8p7 and CEERS-1025, we measure low to moderate \ion{[O}{iii]} and \ion{H}{$\beta$} EWs of ($\EW_{\ion{[O}{iii]}} = {587}_{-28}^{+31}$\,\AA{} and $\EW_{\ion{H}{$\beta$}} = 88_{-13}^{+13}$\,\AA). However, we infer a comparably high ionizing photon production efficiency as we find in EGSY8p7 ($\xi_\text{ion} = 10^{25.6}$\,Hz\,erg$^{-1}$).

Finally, ID 89540 is a similarly bright galaxy as 70289 with $m_\text{F150W} = 26.1$, also previously observed by RUBIES (ID 48045). At its redshift of $z = 8.716$, 89540 lies 1.4\,pMpc away from EGSY8p7 and has an absolute UV magnitude of $\Muv = -21.2$. The faintest of the sample in the rest-optical, we detect \ion{[O}{iii]} and weak \ion{H}{$\beta$}. We do not detect any rest-UV lines in this object and place a $3\sigma$ upper limit on the EW of \Lya, $\EW_{0, \Lya} < 29$\,\AA, though we again note that due to this object being fainter in the rest-UV than the two objects with \Lya{} detections, this upper limit is fully consistent with the lower-EW \Lya{} observed in EGSY8p7 and CEERS-1025. Like 70289, we observe relatively weak \ion{[O}{iii]} and \ion{H}{$\beta$} ($\EW_{\ion{[O}{iii]}} = {523}_{-80}^{+52}$\,\AA{} and $\EW_{\ion{H}{$\beta$}} = 67_{-37}^{+33}$\,\AA) with an ionizing photon production efficiency of $\xi_\text{ion} = 10^{25.5}$\,Hz\,erg$^{-1}$ and a moderately young mass-weighted age of $\sim 35$\,Myr, as inferred from our SED models.

\bigskip

\begin{figure}
    \includegraphics[width=\columnwidth]{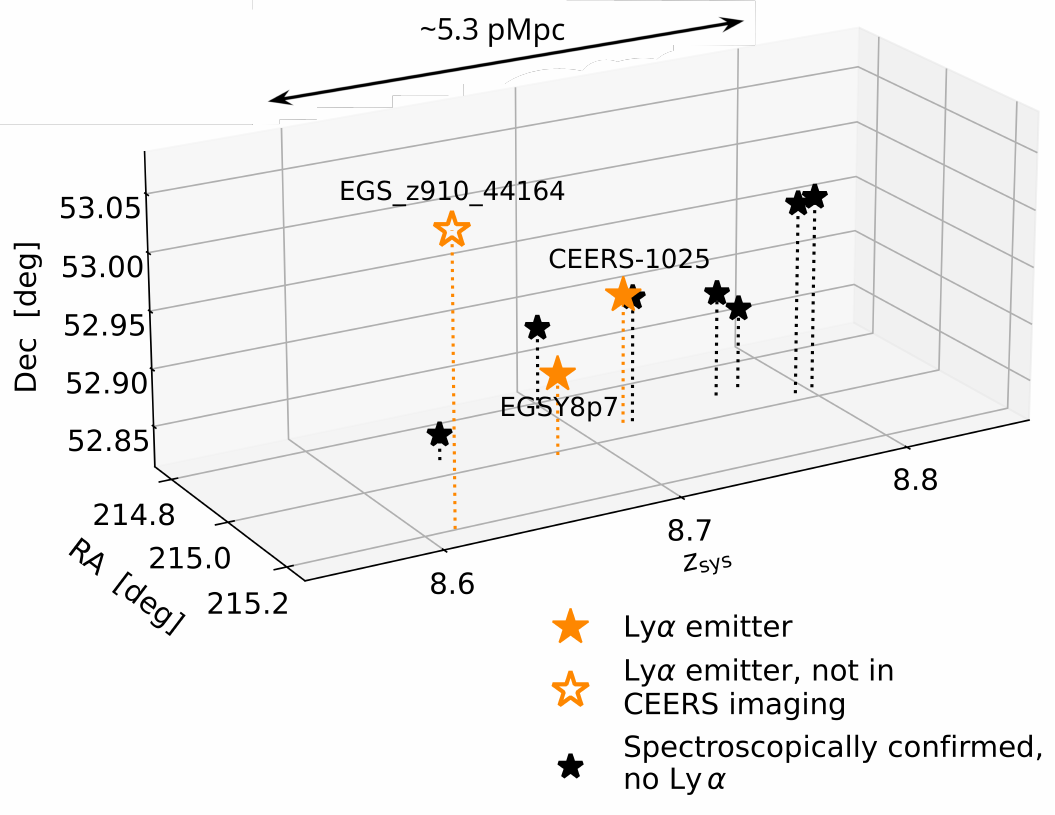}
    \caption{\label{fig:3d_map}The 3D distribution of objects that have been spectroscopically confirmed in the EGS field at $z = 8.7 \pm 0.1$. LAEs are shown as yellow stars and objects that are spectroscopically confirmed but have no \Lya{} detections are shown as black circles. One LAE (EGS\textunderscore z910\textunderscore 44164) has been observed with NIRSpec as part of the CEERS program (ID CEERS-1029) but does not fall in the CEERS imaging footprint and is shown as the open yellow star. There are ten objects with UV luminosities $\Muv \lesssim -19$ that lie within the imaging or in the close vicinity in a narrow redshift range of $\Delta z \sim 0.2$, corresponding to $\sim 5.3$\,pMpc along the line of sight. In comparison, measurements of the $z \sim 9$ UV luminosity function suggest that no more than six objects at $z = 8.6 - 8.8$ and $\Muv \lesssim -19$ are expected in the CEERS imaging area in an average field, hinting at a galaxy overdensity at $z \sim 8.7$ in the volume.}
\end{figure}

Overall, the four $z = 8.7$ objects that we have observed show a variety of properties that may facilitate the growth of a large ionized bubble within the first $\sim 550$ million years after the Big Bang, including \Lya{} emission and signatures of hard ionizing radiation fields (and/or AGN) that may be effective at creating a large ionizing bubble at early times. However, the two LAEs also have indications that they may have strong intrinsic \Lya{} emission, decreasing the need for a large ionized bubble to facilitate the transmission of \Lya{} through the IGM. The properties of these objects do not necessarily imply or require an ionized bubble at early times, but it may nevertheless be possible for a large population of galaxies in this volume to carve out a large ionized region if, for example, they are undergoing rapid bursts of intense star formation and ionizing photon production. Thus, for the remainder of this paper, we investigate the empirical constraints on the presence or lack of a large ionized bubble that are enabled by our observations.

\section{A galaxy overdensity at \texorpdfstring{$\lowercase{z} = 8.7$}{z=8.7}} \label{sec:overdensity}

The EGS field has been observed extensively with targeted JWST multi-object spectroscopy as part of CEERS, RUBIES, CAPERS, and GO 4287. Together, these programs have confirmed the redshifts of ten galaxies at redshifts between $z_\text{sys} = 8.6 - 8.8$ (spanning $\sim 5.3$\,pMpc along the line of sight; we show the 3D distribution of these sources in Figure\ \ref{fig:3d_map}) with UV luminosities ranging from $-22 \lesssim \Muv \lesssim -19$. Of these ten galaxies, nine fall within the $\sim 92$\,arcmin$^2$ CEERS imaging area. In comparison, the UV luminosity function at $z \sim 9$ \citep{Bouwens2021, Donnan2024} implies that the CEERS area is expected to have $\sim 4 - 6$ galaxies of the same luminosity at $z = 8.6 - 8.8$. That is, despite the expected incompleteness of targeted spectroscopic observations, there are still more galaxies confirmed at these redshifts in the EGS field than is expected in an average field, consistent with the presence of a galaxy overdensity in the EGS volume at $z = 8.6 - 8.8$ \citep[also see][]{Finkelstein2022_candels, Larson2022, Whitler2024}.

\begin{figure}
    \includegraphics[width=\columnwidth]{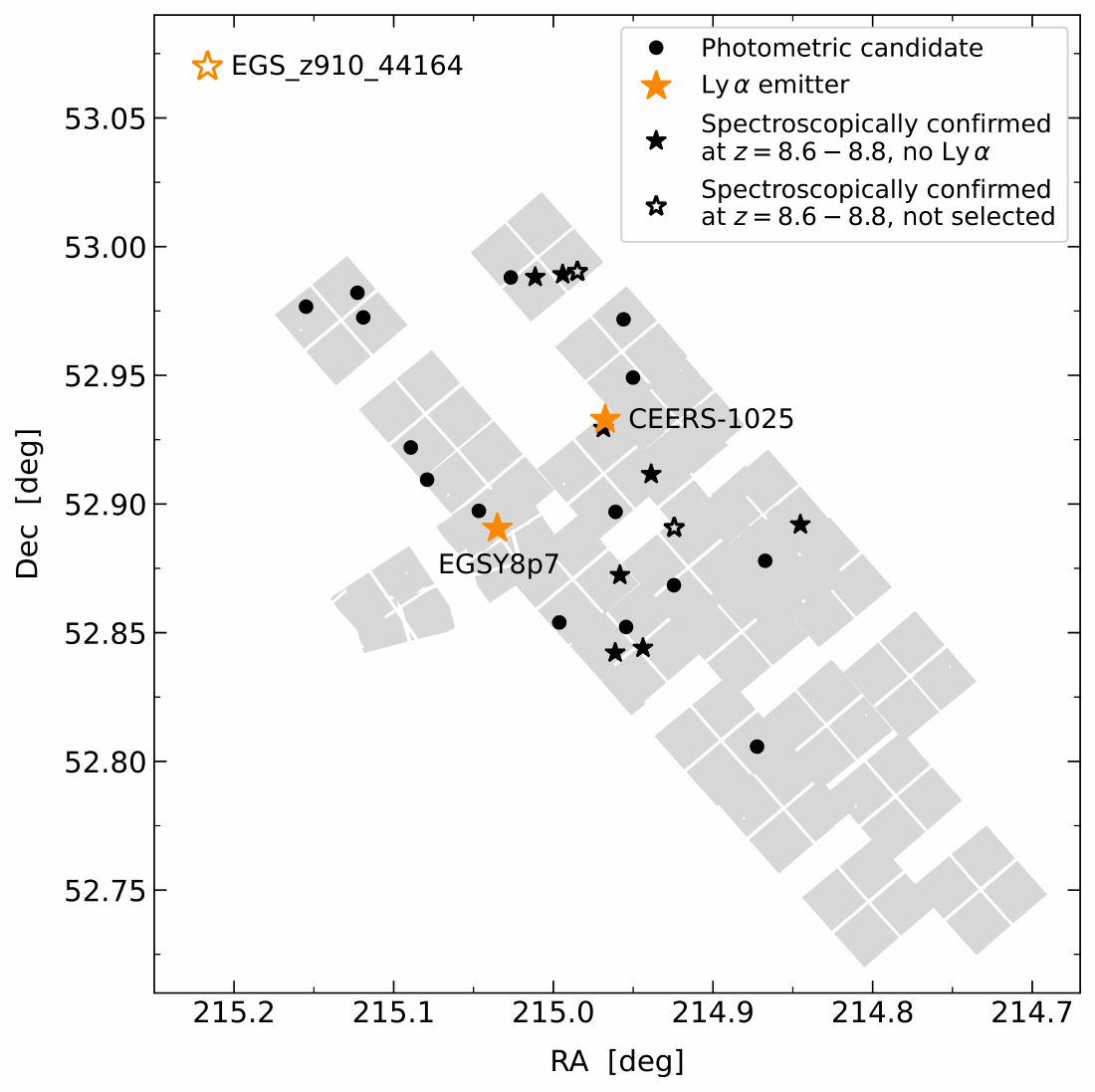}
    \caption{\label{fig:egs_2d_map}The on-sky distribution of our photometric sample. Photometric candidates are shown as black circles, spectroscopically confirmed galaxies without \Lya{} detections are shown as black stars, and LAEs are shown as yellow stars. As in Figure\ \ref{fig:3d_map}, one LAE does not fall in the CEERS imaging footprint and is shown as an open yellow star. Across the entire field, the surface density of this photometric sample is consistent with expectations from the $z \sim 9$ UV luminosity function, but qualitatively, the candidates tend to lie in the northwestern region of the imaging, close to the LAEs, leading to a mild overdensity where an ionized bubble may be expected.}
\end{figure}

\begin{figure*}
    \centering
    \includegraphics[width=\textwidth]{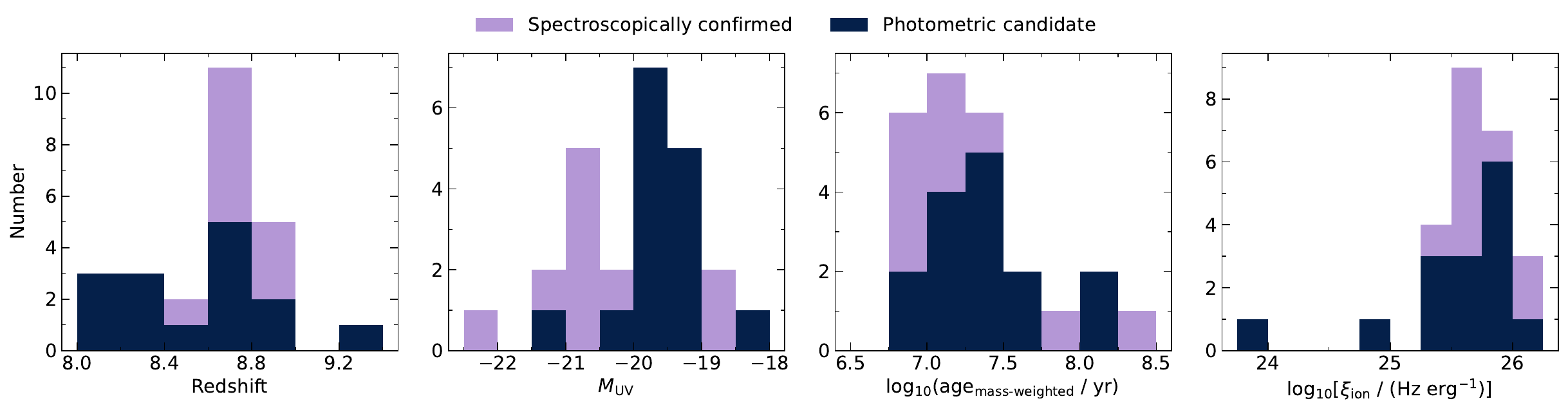}
    \caption{\label{fig:phot_sample_obs_props}The properties of the photometric sample in EGS inferred by our high-redshift \beagle{} SED models (Section\ \ref{sec:overdensity}). We show spectroscopically confirmed objects and photometric candidates as stacked histograms in light purple and dark blue, respectively. From left to right, we show redshift (spectroscopic redshift if available, median photometric redshift if not), absolute UV magnitude, mass-weighted age, and ionizing photon production efficiency, computed using the unattenuated stellar and nebular UV spectrum. We select objects with redshifts ranging from $z = 8.1 - 9.3$ and highlight that the distribution of redshifts peaks at $z = 8.6 - 8.8$, as would be expected from a galaxy overdensity associated with an ionized bubble at $z = 8.7$. The sample ranges from very bright (EGSY8p7 with an absolute UV magnitude of $\Muv = -22.1$ and the next brightest object with $\Muv = -21.3$) to moderately faint (faintest absolute UV magnitude of $\Muv = -18.5$), with a median of $\Muv = -19.8$. The objects in our sample have mass-weighted ages ranging from 6\,Myr to 154\,Myr (median 18\,Myr), and are generally efficient at producing ionizing photons, though three have ionizing photon production efficiencies less than $10^{25}$\,Hz\,erg$^{-1}$ ($\xi_\text{ion}$ ranging between $10^{23.7} - 10^{25.9}$\,Hz\,erg$^{-1}$ for the sample, median $10^{25.5}$\,Hz\,erg$^{-1}$).}
\end{figure*}
To further explore the possible galaxy overdensity in this volume, we update the photometric search for $z \sim 8.7 - 9.1$ galaxy candidates that was conducted by \citet{Whitler2024}. In particular, we incorporate F090W imaging from GO 2234 that enables more robust constraints on nondetections blueward of the \Lya{} break than was possible with the ACS imaging alone that was used by \citet{Whitler2024}. We describe the updated photometric selection criteria and details of the photometric sample in Section\ \ref{subsec:photometric_selection}, then quantify the implied galaxy overdensity in Section\ \ref{subsec:overdensity_calc}.

\subsection{Photometric sample} \label{subsec:photometric_selection}

We select galaxy candidates with a set of color criteria designed to identify partial F115W dropouts at $z \sim 8.7$, similar to the selection used by \citet{Whitler2024}. We note that because we have a new NIRCam filter where we expect $z = 8.7$ galaxies to be undetected, we do not use ACS/F435W data for our primary color selection, as its imaging footprint only partially overlaps with the CEERS footprint. In detail, we adopt the following criteria:
\begin{enumerate}
    \item $\text{S/N} < 3$ in F606W and F814W
    \item $\text{S/N} < 2$ in F090W
    \item $\text{S/N} > 5$ in at least one of F150W and F200W
    \item $\text{S/N}> 3$ in at least two of F277W, F356W, and F444W
    \item $\text{F814W} - \text{F150W} > 1.7$ and $\text{F090W} - \text{F150W} > 1.7$
    \item $\text{F115W} - \text{F150W} > 0.6$ and $\text{F115W} - \text{F150W} < 1.7$
    \item $\text{F150W} - \text{F277W} < 0.6$, and
    \item $\text{F115W} - \text{F150W} > 1.5 \times (\text{F150W} - \text{F277W}) + 0.6$.
\end{enumerate}
Criteria (i) and (ii) are designed to ensure that objects are not detected with significance in filters that are expected to be at shorter wavelengths than the \Lya{} break, while criteria (iii) and (iv) ensure detections in filters at longer wavelengths than the \Lya{} break. Criterion (v) is designed to select objects that fully drop out in F814W and F090W, but crucially, in order to identify $z \sim 8.7$ objects in a relatively narrow redshift range (for a wide band selection), criterion (vi) requires that candidates only partially drop out in F115W. Finally, criteria (vii) and (viii) are designed to reject low-redshift, dusty contaminants by requiring that the observed SEDs of the candidates are only moderately red, while still allowing red objects with a strong \Lya{} break to be selected.

After applying these criteria and performing a visual inspection, which results in the removal of one diffraction spike, we select 39 objects. Of these 39 candidates, ten are spectroscopically confirmed at redshifts between $z = 8.44 - 8.95$. We note that we do not select two spectroscopically confirmed objects at $z = 8.6 - 8.8$: one is not identified in our detection catalog, and the other is not selected due to a formal $\text{S/N} = 2.3$ detection in F090W (though this object satisfies all of our other selection criteria). We use a custom aperture to manually perform photometry at the expected location of the object that is not detected by our standard detection and photometry methods, and find that this object has an $\text{F090W} - \text{F150W}$ color of 1.4, slightly too blue to pass criterion (v) of our selection (though it passes all other criteria).

For the 29 candidates without spectroscopic redshifts, we measure photometric redshifts by modelling the filters expected to probe the rest-UV at $z \sim 8.7$ (ACS F435W, F606W, F814W, and NIRCam F090W, F115W, F200W, and F277W) with \beagle. We adopt a very similar model setup as we used to infer the properties of our spectroscopic sample (Section\ \ref{sec:z8p7_sample}), with the addition of a free redshift parameter ranging uniformly from $z = 0 - 25$. Using the results of these \beagle{} models, we further clean our sample by requiring that candidates have an integrated $z > 8$ probability greater than 90\,per\,cent. We obtain a final sample of 25 objects (ten spectroscopically confirmed galaxies and 15 photometric candidates), distributed across the imaging as shown in Figure\ \ref{fig:egs_2d_map}. These 25 objects range from extremely bright to moderately faint with observed F150W magnitudes of $25.1 \lesssim m_\text{F150W} \lesssim 28.8$ with median $m_\text{F150W} = 27.4$ ($\sim 10 - 320$\,nJy, median $\sim 40$\,nJy). We also measure rest-UV continuum slopes, $\beta$, by fitting a power law ($f_\lambda \propto \lambda^\beta$) to the observed F150W, F200W, and F277W fluxes. We find relatively blue UV slopes ranging from $-2.8 \lesssim \beta \lesssim -1.5$ with a median of $\beta = -2.4$, consistent with expectations for high-redshift galaxies \citep[e.g.][]{Nanayakkara2023, Cullen2023, Cullen2024, Morales2024, Austin2024, Topping2024_uv_slopes, Saxena2024}.

\begin{table*}
    \renewcommand{\arraystretch}{1.4}
    \centering
    \caption{\label{tab:specz_sources}Summary of the spectroscopic observational programs for objects in our photometric sample that have spectroscopic redshifts.}
    \begin{tabular}{cccc}
        \hline
        ID & $z_\text{spec}$ & Obs. Program(s) & Other IDs \\ \hline\hline
        24036 & $8.440$ & RUBIES & RUBIES-18807 \\
        23486 & $8.638$ & CEERS & CEERS-80083 \\
        39259 & $8.677$ & CEERS, RUBIES, GO 4287 & EGSY8p7, CEERS-1019, 69787$^*$ \\
        39700 & $8.687$ & RUBIES, GO 4287 & RUBIES-980841, 70289$^*$ \\
        49539 & $8.716$ & CEERS, GO 4287 & CEERS-1025, 87873$^*$ \\
        50646 & $8.716$ & RUBIES, CAPERS, GO 4287 & RUBIES-48045, CAPERS-11964, 89540$^*$ \\
        51525 & $8.763$ & CEERS (DD 2750) & CEERS-28 \\
        40583 & $8.790$ & CEERS, CAPERS & CEERS-2, CAPERS-4347 \\
        57913 & $8.866$ & CEERS, CAPERS & CEERS-7, CAPERS-6691 \\
        33226 & $8.948$ & CAPERS, GO 4287 & CAPERS-55806 \\
        \hline
    \end{tabular} \\
    $^*$ID in this work and/or \citet{Whitler2024}
\end{table*}

\begin{table*}
    \renewcommand{\arraystretch}{1.4}
    \centering
    \caption{\label{tab:phot_sample_obs_props}Observed and inferred physical properties of the photometric sample, ordered by increasing redshift. We report coordinates, redshifts, apparent magnitudes in F150W ($m_\text{F150W}$), rest-UV continuum slopes ($\beta$), absolute UV magnitudes ($\Muv$), dust extinction at rest-frame 1500\,\AA{} ($A_{1500}$), stellar masses ($M_*$), mass-weighted ages, and ionizing photon production efficiencies ($\xi_\text{ion}$). When possible, we report spectroscopic redshifts without uncertainties (see Table\ \ref{tab:specz_sources} for observational programs). Rest-UV continuum slopes are measured by fitting a power law of the form $f_\lambda \propto \lambda^\beta$ to the observed F150W, F200W, and F277W photometry and ionizing photon production efficiencies are computed considering the intrinsic (i.e. before dust attenuation) stellar and nebular spectrum. We identify EGSY8p7, CEERS-1025, 70289, and 89540 with footnotes, and also report the properties for the two objects with spectroscopic confirmations that were not included in our photometric sample (see Section\ \ref{subsec:photometric_selection}) at the end of the table.}
    \begin{tabular}{ccccccccccc}
        \hline
        \multirow{2}{*}{ID} & RA & Dec & \multirow{2}{*}{Redshift} & \multirow{2}{*}{$m_\text{F150W}$} & \multirow{2}{*}{$\beta$} & \multirow{2}{*}{\Muv} & A$_{1500}$ & \multirow{2}{*}{$\log_{10}\Big(\frac{M_*}{\Msun}\Big)$} & Age & \multirow{2}{*}{$\log_{10}\Big(\frac{\xi_\text{ion}}{\text{Hz } \text{erg}^{-1}}\Big)$} \\
        & [deg] & [deg] & & & & & [mag] & & [Myr] & \\ \hline\hline
        \multicolumn{10}{c}{Photometric sample} \\ \hline
        55861 & 215.04704 & +52.89748 & $8.10_{-0.10}^{+0.07}$ & $27.5_{-0.1}^{+0.1}$ & $-2.1 \pm 0.1$ & $-19.5_{-0.1}^{+0.1}$ & $0.48_{-0.13}^{+0.14}$ & $7.6_{-0.5}^{+0.5}$ & $7_{-4}^{+58}$ & $25.69_{-0.07}^{+0.05}$ \\
        44302 & 214.95007 & +52.94927 & $8.15_{-0.14}^{+0.19}$ & $27.7_{-0.1}^{+0.1}$ & $-2.2 \pm 0.2$ & $-19.3_{-0.1}^{+0.1}$ & $0.19_{-0.12}^{+0.17}$ & $7.4_{-0.4}^{+0.5}$ & $9_{-7}^{+74}$ & $25.77_{-0.08}^{+0.07}$ \\
        52577 & 215.08994 & +52.92206 & $8.20_{-0.14}^{+0.19}$ & $27.2_{-0.1}^{+0.1}$ & $-2.1 \pm 0.1$ & $-19.9_{-0.1}^{+0.1}$ & $0.31_{-0.21}^{+0.33}$ & $8.6_{-0.9}^{+0.4}$ & $142_{-138}^{+193}$ & $25.86_{-0.09}^{+0.05}$ \\
        57935 & 215.12373 & +52.98215 & $8.34_{-1.58}^{+0.69}$ & $27.8_{-0.1}^{+0.1}$ & $-1.5 \pm 0.2$ & $-19.1_{-0.4}^{+0.9}$ & $2.21_{-0.69}^{+0.85}$ & $8.8_{-0.3}^{+0.2}$ & $16_{-4}^{+7}$ & $24.92_{-0.29}^{+0.29}$ \\
        41077 & 214.95602 & +52.97191 & $8.38_{-0.19}^{+0.20}$ & $27.8_{-0.1}^{+0.1}$ & $-2.0 \pm 0.2$ & $-19.4_{-0.1}^{+0.1}$ & $0.28_{-0.16}^{+0.21}$ & $7.7_{-0.4}^{+0.4}$ & $32_{-31}^{+102}$ & $25.81_{-0.05}^{+0.05}$ \\
        45345 & 214.96104 & +52.89712 & $8.38_{-0.14}^{+0.14}$ & $27.7_{-0.1}^{+0.1}$ & $-2.4 \pm 0.2$ & $-19.6_{-0.1}^{+0.1}$ & $0.10_{-0.07}^{+0.16}$ & $7.9_{-0.7}^{+0.4}$ & $19_{-12}^{+52}$ & $25.53_{-0.18}^{+0.12}$ \\
        24036 & 214.94383 & +52.84423 & $8.440$ & $26.6_{-0.1}^{+0.1}$ & $-2.4 \pm 0.1$ & $-20.7_{-0.0}^{+0.0}$ & $0.11_{-0.06}^{+0.09}$ & $8.3_{-0.3}^{+0.3}$ & $17_{-8}^{+21}$ & $25.51_{-0.09}^{+0.09}$ \\
        31896 & 214.92435 & +52.86860 & $8.58_{-0.21}^{+0.17}$ & $28.1_{-0.1}^{+0.1}$ & $-2.5 \pm 0.3$ & $-19.2_{-0.1}^{+0.1}$ & $0.21_{-0.16}^{+0.39}$ & $8.3_{-0.5}^{+0.3}$ & $46_{-30}^{+60}$ & $25.19_{-0.65}^{+0.27}$ \\
        57213 & 215.12004 & +52.97256 & $8.61_{-0.10}^{+0.09}$ & $27.0_{-0.1}^{+0.1}$ & $-2.5 \pm 0.1$ & $-20.3_{-0.0}^{+0.1}$ & $0.09_{-0.06}^{+0.10}$ & $8.2_{-0.7}^{+0.6}$ & $21_{-14}^{+52}$ & $25.47_{-0.23}^{+0.15}$ \\
        56931 & 215.15613 & +52.97668 & $8.63_{-0.20}^{+0.15}$ & $27.7_{-0.1}^{+0.1}$ & $-2.2 \pm 0.2$ & $-19.6_{-0.1}^{+0.2}$ & $0.39_{-0.33}^{+0.60}$ & $8.4_{-0.4}^{+0.4}$ & $25_{-11}^{+27}$ & $25.10_{-0.61}^{+0.32}$ \\
        27069 & 214.99640 & +52.85415 & $8.63_{-0.13}^{+0.12}$ & $27.3_{-0.1}^{+0.1}$ & $-2.6 \pm 0.2$ & $-19.8_{-0.1}^{+0.1}$ & $0.07_{-0.05}^{+0.12}$ & $8.0_{-0.6}^{+0.4}$ & $16_{-10}^{+39}$ & $25.49_{-0.17}^{+0.15}$ \\
        23486 & 214.96129 & +52.84236 & $8.638$ & $28.1_{-0.1}^{+0.1}$ & $-1.8 \pm 0.2$ & $-19.0_{-0.1}^{+0.1}$ & $0.43_{-0.19}^{+0.21}$ & $7.3_{-0.4}^{+0.4}$ & $8_{-7}^{+66}$ & $25.80_{-0.07}^{+0.06}$ \\
        50620 & 215.07962 & +52.90956 & $8.65_{-0.05}^{+0.05}$ & $27.4_{-0.1}^{+0.1}$ & $-2.4 \pm 0.1$ & $-19.8_{-0.1}^{+0.1}$ & $0.05_{-0.03}^{+0.06}$ & $7.6_{-0.4}^{+0.3}$ & $16_{-11}^{+58}$ & $25.76_{-0.18}^{+0.04}$ \\
        39259$^a$ & 215.03539 & +52.89067 & $8.677$ & $25.1_{-0.0}^{+0.0}$ & $-1.8 \pm 0.0$ & $-22.1_{-0.0}^{+0.0}$ & $0.13_{-0.07}^{+0.08}$ & $7.9_{-0.5}^{+0.3}$ & $16_{-15}^{+56}$ & $25.78_{-0.04}^{+0.03}$ \\
        39700$^b$ & 214.84477 & +52.89211 & $8.687$ & $26.6_{-0.0}^{+0.0}$ & $-1.8 \pm 0.1$ & $-20.6_{-0.0}^{+0.0}$ & $1.51_{-0.78}^{+0.67}$ & $9.1_{-0.4}^{+0.3}$ & $154_{-116}^{+231}$ & $25.58_{-0.20}^{+0.15}$ \\
        26571 & 214.95444 & +52.85239 & $8.70_{-0.24}^{+0.18}$ & $28.8_{-0.1}^{+0.2}$ & $-1.5 \pm 0.3$ & $-18.5_{-0.1}^{+0.2}$ & $1.03_{-0.33}^{+0.32}$ & $7.8_{-0.6}^{+0.5}$ & $18_{-13}^{+56}$ & $25.62_{-0.21}^{+0.13}$ \\
        49539$^c$ & 214.96753 & +52.93296 & $8.716$ & $26.1_{-0.0}^{+0.0}$ & $-2.4 \pm 0.0$ & $-21.3_{-0.0}^{+0.0}$ & $0.09_{-0.05}^{+0.08}$ & $7.9_{-0.1}^{+0.1}$ & $13_{-2}^{+4}$ & $25.35_{-0.08}^{+0.10}$ \\
        50646$^d$ & 214.96869 & +52.92965 & $8.716$ & $26.5_{-0.1}^{+0.1}$ & $-2.5 \pm 0.1$ & $-20.8_{-0.0}^{+0.0}$ & $0.65_{-0.31}^{+0.43}$ & $8.1_{-0.5}^{+0.4}$ & $32_{-21}^{+70}$ & $25.46_{-0.13}^{+0.17}$ \\
        51525 & 214.93863 & +52.91175 & $8.763$ & $26.4_{-0.0}^{+0.0}$ & $-2.4 \pm 0.1$ & $-20.9_{-0.0}^{+0.0}$ & $0.11_{-0.07}^{+0.12}$ & $7.8_{-0.5}^{+0.5}$ & $6_{-3}^{+30}$ & $25.66_{-0.09}^{+0.06}$ \\
        40583 & 214.99440 & +52.98938 & $8.790$ & $27.0_{-0.0}^{+0.0}$ & $-2.5 \pm 0.1$ & $-20.3_{-0.0}^{+0.0}$ & $0.08_{-0.06}^{+0.15}$ & $8.3_{-0.2}^{+0.3}$ & $23_{-11}^{+25}$ & $25.47_{-0.13}^{+0.10}$ \\
        57913 & 215.01170 & +52.98831 & $8.866$ & $26.6_{-0.0}^{+0.0}$ & $-2.5 \pm 0.1$ & $-20.7_{-0.0}^{+0.0}$ & $0.08_{-0.07}^{+0.20}$ & $8.4_{-0.1}^{+0.1}$ & $15_{-4}^{+10}$ & $25.45_{-0.17}^{+0.09}$ \\
        35351 & 214.86694 & +52.87808 & $8.91_{-0.16}^{+0.14}$ & $27.7_{-0.1}^{+0.1}$ & $-2.8 \pm 0.2$ & $-19.4_{-0.1}^{+0.1}$ & $0.04_{-0.03}^{+0.10}$ & $7.9_{-0.2}^{+0.3}$ & $20_{-8}^{+20}$ & $25.33_{-0.37}^{+0.15}$ \\
        33226 & 214.95833 & +52.87252 & $8.948$ & $28.7_{-0.1}^{+0.2}$ & $-2.3 \pm 0.3$ & $-18.6_{-0.1}^{+0.1}$ & $0.34_{-0.19}^{+0.29}$ & $8.0_{-0.4}^{+0.4}$ & $31_{-19}^{+57}$ & $25.35_{-0.16}^{+0.18}$ \\
        12264 & 214.87219 & +52.80588 & $8.96_{-0.04}^{+0.04}$ & $26.3_{-0.0}^{+0.0}$ & $-1.9 \pm 0.1$ & $-21.2_{-0.0}^{+0.0}$ & $0.44_{-0.07}^{+0.07}$ & $9.0_{-0.0}^{+0.0}$ & $16_{-1}^{+1}$ & $24.68_{-0.09}^{+0.09}$ \\
        42161 & 215.02708 & +52.98818 & $9.27_{-0.16}^{+0.17}$ & $27.4_{-0.1}^{+0.1}$ & $-2.5 \pm 0.2$ & $-19.9_{-0.1}^{+0.1}$ & $0.05_{-0.04}^{+0.10}$ & $9.2_{-0.1}^{+0.1}$ & $104_{-22}^{+36}$ & $23.73_{-0.10}^{+0.40}$ \\ \hline
        \multicolumn{10}{c}{Spectroscopically confirmed, not in photometric sample} \\ \hline
        RUBIES- & \multirow{2}{*}{214.92415} & \multirow{2}{*}{+52.89096} & \multirow{2}{*}{8.774} & \multirow{2}{*}{$28.1_{-0.15}^{+0.18}$} & \multirow{2}{*}{$-1.9 \pm 0.3$} &  \multirow{2}{*}{$-19.2_{-0.1}^{+0.1}$} & \multirow{2}{*}{$0.11_{-0.10}^{+0.27}$} & \multirow{2}{*}{$7.6_{-0.6}^{+0.6}$} & \multirow{2}{*}{$12_{-9}^{+71}$} & \multirow{2}{*}{$25.92_{-0.14}^{+0.14}$} \\
        45438 & & & & & & & & & \\
        CAPERS- & \multirow{2}{*}{214.98511} & \multirow{2}{*}{+52.99047} & \multirow{2}{*}{8.799} & \multirow{2}{*}{$26.6_{-0.12}^{+0.06}$} & \multirow{2}{*}{$-2.2 \pm 0.1$} & \multirow{2}{*}{$-20.7_{-0.1}^{+0.1}$} & \multirow{2}{*}{$0.12_{-0.10}^{+0.21}$} & \multirow{2}{*}{$8.6_{-0.9}^{+0.4}$} & \multirow{2}{*}{$32_{-27}^{+49}$} & \multirow{2}{*}{$25.52_{-0.25}^{+0.23}$} \\
        3044 & & & & & & & & & \\ \hline
    \end{tabular}
    $^a$EGSY8p7; $^b$70289; $^c$CEERS-1025; $^d$89540
\end{table*}

To infer the physical properties of these 25 objects, we re-fit their observed SEDs with \beagle{} models that consider all of their available photometric data. We also restrict the uniform redshift prior to $z = 6 - 10$, as we assume that our previous models with a free redshift parameter have identified objects at low redshift. Otherwise, we keep the same model parameters. We show the distributions of redshifts, \Muv{}, mass-weighted ages, and ionizing photon production efficiencies ($\xi_\text{ion}$) before the UV flux has been processed through gas and attenuated by dust in the ISM in Figure\ \ref{fig:phot_sample_obs_props} and report the inferred properties of the entire sample in Table\ \ref{tab:phot_sample_obs_props}, noting that we carry over the SED model results from Section\ \ref{sec:z8p7_sample} for the four objects already discussed. In brief, we infer photometric redshifts between $z = 8.1 - 9.3$ and absolute magnitudes of $\Muv = -22.1$ for EGSY8p7 and $-21.3 \leq \Muv \leq -18.5$ for the rest of the sample, with a median of $\Muv = -19.8$ for the entire sample. We infer that these objects have high ionizing photon production efficiencies (median $\xi_\text{ion} = 10^{25.5}$\,Hz\,erg$^{-1}$), though three have values less than $10^{25}$\,Hz\,erg$^{-1}$ (full range of $\xi_\text{ion} = 10^{23.7} - 10^{25.9}$\,Hz\,erg$^{-1}$). However, we find that the typical mass-weighted age is only a few tens of Myr (median 18\,Myr, though the sample ranges from ages as young as 6\,Myr to as old as 154\,Myr).

Given constraints on these physical properties, we can briefly assess whether the observed galaxies are sufficient to create a large, $R \sim 2$\,pMpc ionized bubble. To this end, we follow the methods of \citet{Whitler2024} (their equation 1) to estimate the radius of a spherical \ion{H}{ii} region that a galaxy population (described by a UV luminosity function) with given ionizing photon production efficiencies and escape fractions could create, if the galaxies were producing ionizing photons at a constant rate over their entire lifetimes. We adopt the redshift evolution of the Schechter parameters found by \citet{Bouwens2021} evaluated at $z = 8.7$ as our UV luminosity function (integrated to $\Muv = -18.5$ for general consistency with the observed UV magnitudes of our sample), fix $\xi_\text{ion}$ to the median value of $10^{25.5}$\,Hz\,erg$^{-1}$ observed for our photometric sample, and assume that the population has been producing ionizing photons constantly for $\sim 20$\,Myr, consistent with the median mass-weighted age of the sample. We adopt a fixed escape fraction of $f_\text{esc} = 0.1$, broadly consistent with the escape fraction predicted by the $\beta - f_\text{esc}$ relation found by \citet{Chisholm2022} for the median UV slope of our sample. These parameters regulate the photoionization rate due to ionizing flux from galaxies in equation 1 of \citet{Whitler2024}. For the recombination rate, we assume a fixed value for the `clumping factor,' which quantifies inhomogeneities in the IGM \citep{Madau1999}, of $C = 3$ \citep[e.g.][]{Shull2012, Finlator2012, Gorce2018}.

Under these assumptions, we find that the radius of the ionized bubble that can be created by the observed population is $\sim 0.3$\,pMpc. This suggests either that there is no extremely large, $R_b \gtrsim 2$\,pMpc ionized bubble that contains both EGSY8p7 and EGS\textunderscore z910\textunderscore 44164, or the observed population was not producing ionizing photons long enough to reionize such a large volume. In the latter case, other, unseen sources of ionizing flux would be required. For example, there may be large numbers of galaxies that are fainter than our observational detection limit, as is implied by UV luminosity functions measured in lensed fields at $z \gtrsim 9$ \citep[e.g.][]{Chemerynska2026}, which may be contributing significantly to the ionizing photon budget. We test this by integrating our assumed luminosity function to the commonly adopted faint limit of $\Muv = -13$ \citep{Robertson2015} rather than $\Muv = -18.5$, and find an ionized bubble radius of $R_b \sim 0.6$, still markedly smaller than $R_b \sim 2$\,pMpc. Alternatively, or additionally, the creation of a large ionized bubble may have also been facilitated by earlier periods of ionized photon production, potentially during bursts of star formation. If the currently observable population traces a large population of bursty galaxies, various members of such a population may have also been extremely star-forming in the past, therefore contributing successive episodes of ionizing photon production towards carving out a large bubble. This process would be additionally aided if the galaxy population in the volume was more abundant than the average.

\subsection{Quantifying the overdensity} \label{subsec:overdensity_calc}

We now investigate the strength of the galaxy overdensity and therefore the population of galaxies that could be reionizing the IGM in this field. We compute the surface density of our sample, corrected for the incompleteness of our photometric selection, then compare to the average surface density expected at these redshifts implied by the UV luminosity function at $z \sim 9$. We quantify our selection function by performing source injection and recovery simulations, where we place mock sources with known properties into the real mosaics, then perform detection, calculate photometry, and do selection using the same methods that we use to assemble the real sample.

To generate the fluxes of the sources that we inject, we define a grid of redshifts ($7 \leq z \leq 12$ with a step of $\Delta z = 0.1$) and absolute UV magnitudes ($-24 \leq \Muv \leq -16$, $\Delta \Muv = 0.2$). We generate mock SEDs for each pair of $(z, \Muv)$, assuming that the SEDs are power laws with rest-UV slopes, $\beta$, determined from the F115W dropout $\beta - \Muv$ relation found by \citet{Topping2024_uv_slopes}, then normalize to \Muv{} at rest-frame 1500\,\AA. We then redshift each mock SED, apply the IGM attenuation model of \citet{Inoue2014}, and calculate `true' photometry in all of the the ACS and NIRCam filters we use in this work.

To create the images of the mock sources to inject into the mosaics, we also generate morphological parameters. We assume S\'ersic surface brightness profiles and sample S\'ersic indices from a one-sided truncated normal distribution with a mean of $\mu_n = 1$, standard deviation of $\sigma_n = 1$, and minimum of $a_n = 0.5$. We sample ellipticities ($e = 1 - b / a$, where $a$ and $b$ are the semi-major and semi-minor axes, respectively, such that a circle has $e = 0$) from a truncated normal distribution with mean $\mu_e = 0.2$, standard deviation $\sigma_e = 0.4$, minimum $a_e = 0$, and maximum $b_e = 1$, and sample position angles between $-90^\circ$ and $+90^\circ$ from a uniform distribution. Finally, we set the sizes of objects in each bin of $(z, \Muv)$ using the \citet{Shibuya2015} size-luminosity relation at the appropriate redshift.

Given these fluxes and morphological parameters, we create mock source images that are normalized such that the total flux of the image in a given filter sums to the flux in that filter, then convolve the image with the appropriate PSF. We then randomly sample positions across the real image and add the normalized mock source images to the mosaics. We create 100 mock sources for each bin of $(z, \Muv)$, which results in a total of $\sim 210\,000$ objects in the mock source catalog. To ensure that these sources are not placed at such high densities that they regularly obstruct one another and artificially lower our completeness, we create five realizations of the mosaics with mock sources, each of which has $\sim 42\,000$ objects (corresponding to a surface density of $\sim 0.1$\,arcsec$^{-2}$). For each of these instances of the mosaics with mock sources, we run detection and calculate photometry, then apply the same selection criteria that we used to select the real sample to the resulting photometric catalogs.

\begin{figure}
    \centering
    \includegraphics[width=\columnwidth]{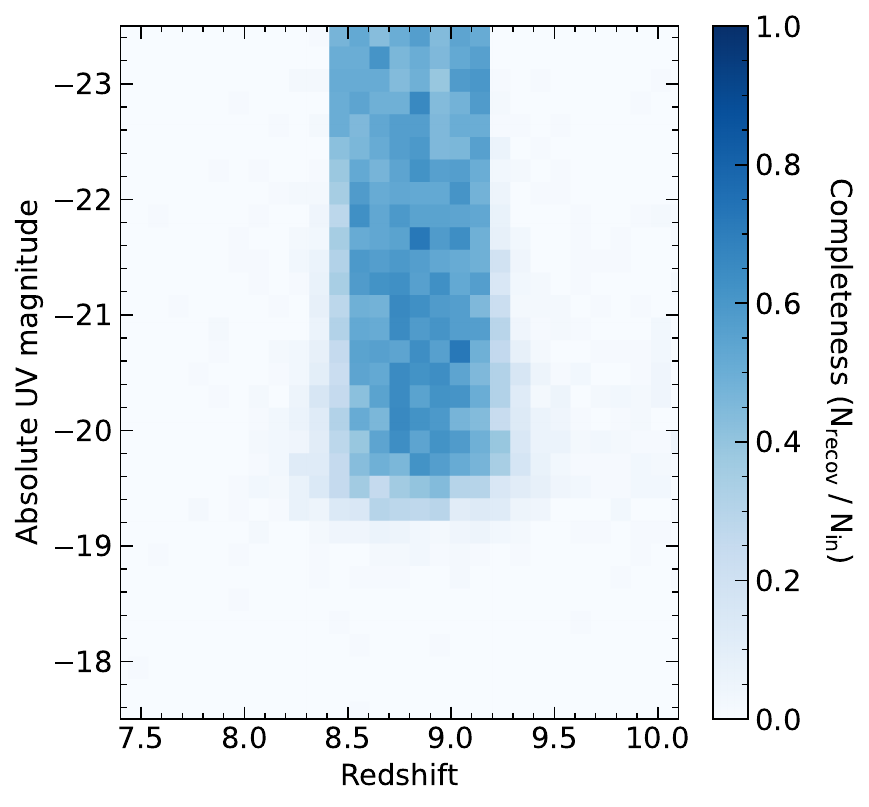}
    \caption{\label{fig:selection_function}Completeness of the photometric selection we use to investigate the photometric galaxy overdensity at $z \sim 8.7$. The selection is most efficient at redshifts of $z \sim 8.5 - 9.2$ and absolute UV magnitudes brighter than $\Muv \sim -19.2$. At these redshifts and luminosities, we find a maximum completeness of $\sim 60$\,per\,cent due to injected objects overlapping with real objects in the image (quantifying the fraction of the imaging area that is impacted by projection effects) and photometric scatter that causes objects to be detected in filters where we expect nondetections (F606W, F814W, and F090W).}
\end{figure}

Finally, we quantify the completeness of our selection as a function of redshift and UV luminosity by calculating $N_\text{recov} / N_\text{in}$ in each bin of $(z, \Muv)$. $N_\text{recov}$ is the total number of sources that were recovered (i.e. detected and selected) over all five realizations of the mosaics in the $(z, \Muv)$ bin under consideration and $N_\text{in}$ is the total number of objects that were originally injected in the same bin. The selection function that results from this calculation is shown in Figure\ \ref{fig:selection_function}; we find that our selection is most sensitive to objects with absolute UV magnitudes of $\Muv \lesssim -19.2$ at redshifts of $z \sim 8.5 - 9.2$, with a maximum completeness of $\sim 60$\,per\,cent due to photometric noise causing formal $\text{S/N} > 2$ or $\text{S/N} > 3$ detections in filters that are expected to be at shorter wavelengths than the \Lya{} break.

Applying this selection function to multiple measurements of the $z \sim 9$ luminosity function (\citealt{Finkelstein2023, Donnan2024}, and the evolutionary model for the Schechter parameters found by \citealt{Bouwens2021}) suggests that the expected, average surface density of galaxies that would be identified by our photometric selection in the area we use in this work is $\sim 0.11 - 0.16$\,arcmin$^{-2}$, such that $10 - 15$ galaxies are expected in the field. In comparison, we have photometrically identified 25 candidates (10 of which have been spectroscopically confirmed) across the entire imaging area in EGS, implying a factor of $\sim 1.7 - 2.5$ overdensity of $\Muv \lesssim -19.2$ galaxies at $z \sim 8.5 - 9.2$ in the area. Furthermore, we note that the on-sky distribution of our sample qualitatively suggests that the density of $z \sim 8.7$ galaxies is highest in the northeastern region of the imaging (see Figure\ \ref{fig:egs_2d_map}), where a bubble may be expected given the proximity to the known LAEs, and therefore where an overdensity may be expected. Thus, we also quantify the surface density as a function of projected distance from the brightest LAE in the field, EGSY8p7, and find a mild overdensity by a factor of $\sim 2.4 - 3.6$ (depending on the UV luminosity function model adopted) in the 5\,arcmin$^2$ closest to EGSY8p7. The surface density then decreases as a function of increasing distance from EGSY8p7 until the observed sample is underdense by a factor of $\sim 1.7 - 2.5$ at separations $\gtrsim 15$\,arcmin$^2$.

This overdensity, especially as it is in the region closest to EGSY8p7, suggests that there may be a slightly larger galaxy population in EGS at $z = 8.7$ than the average, which may be capable of creating a large ionized bubble. Indeed, at lower redshifts ($z \sim 7 -  8.5$), comparably overdense regions have been demonstrated to host galaxies with strong \Lya{} emission, which are expected to trace ionized bubbles \citep{Chen2025}. However, given the moderately young mass-weighted ages of the currently observable $z = 8.7$ sample in EGS, this population is unlikely to be capable of ionizing a very large region alone. Thus, in the next section, we use \Lya{} to directly assess the probability of the existence of a very large, $R_b \gtrsim 2$\,pMpc ionized region that contains both of the two bright LAEs, EGSY8p7 and EGS\textunderscore z910\textunderscore 44164, at $z = 8.7$.

\section{\texorpdfstring{L\lowercase{y}\,$\alpha$}{Lya} transmission through the IGM} \label{sec:lya_transmission}

For galaxies in large ionized bubbles, the transmission of \Lya{} emission through the IGM (\TIGM) is expected to be higher than the average in the field, as their \Lya{} photons can cosmologically redshift into the damping wing while inside the bubble before being attenuated by intergalactic neutral hydrogen. In this section, we quantify the \Lya{} transmission implied by our observations of \Lya{} at $z = 8.7$, then compare with measurements of \TIGM{} in the field \citep{Tang2024_highz_lya} to investigate whether this volume has systematically enhanced \Lya{} transmission indicative of a very large, $R_b \gtrsim 2$\,pMpc ionized bubble. We note that, in this work, we primarily focus on the area near EGSY8p7, as current imaging and spectroscopic datasets only sparsely cover the area between EGSY8p7 and EGS\textunderscore z910\textunderscore 44164. We provide a brief overview of the method we use to measure \TIGM{} \citep[which is, in turn, based on similar methods from the literature;][]{Mason2018, Mason2019_kmos}, but refer to \citet{Tang2024_highz_lya} for a detailed description.

We begin by creating a forward model of the \Lya{} rest-frame EW distribution as a function of \TIGM, which compares the unattenuated \Lya{} EW distribution to the observed \Lya{} rest-frame EWs at $z = 8.7$ that have been attenuated by the IGM. For the unattenuated \Lya{} EW distribution that serves as a baseline, we adopt the average \Lya{} EW distribution for Lyman-break galaxies at $z \sim 5$ found by \citet{Tang2024_z5_lya}, which includes both \Lya{} detections and upper limits, and is modelled as a log-normal distribution\footnote{$p(\EW) = \dfrac{\exp\left[-(\ln(\EW / \text{\AA}) - \mu)^2/{2\sigma^2}\right]}{\EW \sqrt{2\pi\sigma^2}}$} with a mean of $\mu = 2.38$ and a standard deviation of $\sigma = 1.64$. We adopt the $z \sim 5$ EW distribution for two primary reasons. First, reionization is expected to end by $z \sim 5$ \citep[e.g.][]{Yang2020, Zhu2021, Zhu2023, Bosman2022}, so the dominant red peak of \Lya{} is not expected to be impacted by intergalactic \ion{H}{i} at $z \sim 5$, and the $z \sim 5$ \Lya{} EW distribution is expected to capture the distribution without the impact of IGM attenuation. Second, the intrinsic galaxy properties that likely also regulate \Lya{} emission are expected to be more similar to our $z \sim 9$ objects at $z \sim 5$ than at lower redshifts, though future measurements of the \Lya{} EW distribution for subsets of $z \sim 5$ galaxies with properties matched to $z \sim 9$ galaxies will further improve upon this assumption.

We emphasize that the parameters of the $z \sim 5$ \Lya{} EW distribution were derived after correcting for the loss of \Lya{} flux from the NIRSpec MSA aperture. \Lya{} emission in galaxies has been demonstrated to be extended over large halos \citep[e.g.][]{Steidel2011, Leclercq2017_muse_hubble, Leclercq2020_muse_hubble, SaldanaLopez2026_ly_continuum}, implying that small-aperture observations may miss diffuse \Lya{} flux. We refer to \citet{Tang2024_z5_lya, Tang2024_highz_lya} for a detailed description of the NIRSpec MSA aperture correction, but in brief, the correction is based on a model for the surface brightness of \Lya{} informed by the \Lya{} halo measurements of \citet{Leclercq2017_muse_hubble}. The fraction of the integrated flux contained in the NIRSpec aperture is then used to correct the parameters of the EW distribution, which was originally derived from VLT/MUSE integral field spectroscopy and Keck/DEIMOS slit spectroscopy, to reflect NIRSpec shutters. We note that applying this correction to our $z = 8.7$ sample implicitly assumes that the properties of \Lya{} halos do not significantly evolve between $z = 5$ and $z = 8.7$. However, \Lya{} halo properties are not well-constrained at $ z \gtrsim 7$, motivating the need for integral field spectroscopy or other observations such as slit-stepping with the MSA \citep{Barisic2025} to fully characterize the spatial distribution of \Lya{} at the high redshifts of this work.

As we are interested in the transmission at $z = 8.7$ relative to $z \sim 5$, our forward model for the \Lya{} EW distribution depends on \TIGM{} as $p(\EW\,\vert\, \TIGM) = p_{z \sim 5}\left(\frac{\EW}{\TIGM}\right)$ and the transmission at $z \sim 5$ is unity. Then, the likelihood of detecting \Lya{} emission from the $i^\text{th}$ galaxy with a rest-frame \Lya{} equivalent width of $\EW_i$ and $1\sigma$ uncertainty of $\sigma_i$ is
\begin{equation} \label{eqn:lya_detection_likelihood}
    p(\EW_i \,\vert\, \TIGM)_\text{det} = \int_0^\infty p(\EW\,\vert\, \TIGM) \frac{\exp\left[-\frac{(\EW - \EW_i)^2}{2\sigma_i^2}\right]}{\sqrt{2\pi}\sigma_i} \, \text{d}\EW.
\end{equation}
For a non-detection of \Lya{} with a $3\sigma$ upper limit on the EW of $\EW_{3\sigma, i}$ and $1\sigma$ uncertainty of $\sigma_i = \EW_{3\sigma, i} / 3$, the likelihood is
\begin{equation} \label{eqn:lya_nondetection_likelihood}
    p(\EW_i \,\vert\, \TIGM)_\text{nondet} = \int_0^\infty p(\EW\,\vert\, \TIGM) \frac{\text{erfc}\left(\frac{\EW - \EW_{3\sigma, i}}{\sqrt{2}\sigma_i}\right)}{2} \, \text{d}\EW.
\end{equation}
By Bayes' Theorem, the posterior probability distribution for the IGM transmission in EGS at $z = 8.7$ given the set of observed \Lya{} EWs, $p(\TIGM \,\vert\,\{\EW_i\})$, is then
\begin{equation}
p(\TIGM \,\vert\,\{\EW_i\}) \propto p(\TIGM) \cdot \prod_i p(\EW_i\,\vert\, \TIGM),
\end{equation}
where $p(\EW_i\,\vert\, \TIGM)$ is the likelihood for each individual galaxy with a \Lya{} detection (Equation\ \ref{eqn:lya_detection_likelihood}) or non-detection (Equation\ \ref{eqn:lya_nondetection_likelihood}) and $p(\TIGM)$ is the prior on \TIGM, which we take to be uniform between zero and one.

\begin{figure}
    \includegraphics[width=\columnwidth]{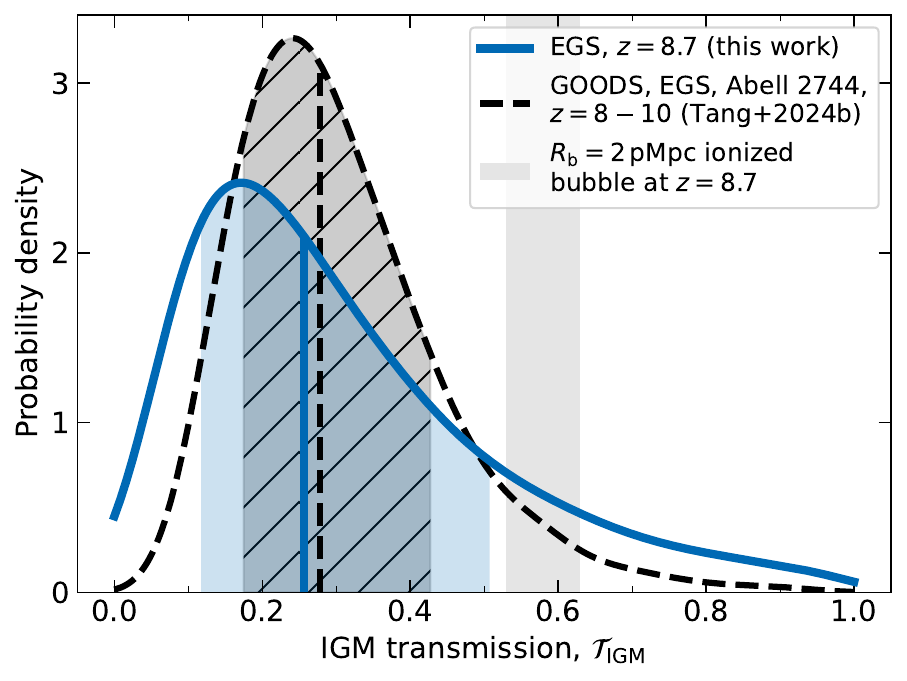}
    \caption{\label{fig:T_IGM}The posterior probability distribution for the transmission of \Lya{} through the IGM, \TIGM, at $z = 8.7$ in EGS. We show our measurement of \TIGM{} in EGS at $z = 8.7$, inferred by combining our four \Lya{} constraints with \Lya{} constraints from the literature (solid blue line and shaded region) compared with the constraint on \TIGM{} in the field at $z = 8 - 10$ found by \citet{Tang2024_highz_lya} (dashed black line and hatched shaded region). We also show the the expectation for the transmission of \Lya{} through a very large ionized bubble with radius $R_b = 2$\,pMpc with and without attenuation by residual neutral hydrogen in infalling gas (grey shaded region). Our constraint on the transmission of \Lya{} is fully consistent with the field measurement, which is in turn slightly less than the transmission expected in the presence of an $R_b = 2$\,pMpc ionized bubble. This tentatively suggests the \Lya{} emitters in this volume are not embedded in such an ionized region, which would be unexpected at this redshift, but it may still be possible for these objects to inhabit a more typical, slightly smaller (e.g. $R_b \sim 0.5 - 1$\,pMpc) ionized bubble.}
\end{figure}

We sample $p(\TIGM \,\vert\,\{\EW_i\})$ with a Markov Chain Monte Carlo (MCMC) algorithm using the \texttt{emcee} \citep{Foreman-Mackey2013} package. We run the MCMC for 10\,000 steps using 32 walkers, initializing each walker by sampling from a Gaussian distribution centered at $\TIGM = 0.5$ with a standard deviation of $\sigma_{\TIGM} = 0.1$. We do not use a burn-in period, and thus produce a total of 320\,000 samples with equal weights. We concatenate the chains from all walkers to obtain the full posterior for \TIGM{} at $z = 8.7$ in EGS, shown in Figure\ \ref{fig:T_IGM}, then take the median, $16^\text{th}$ and $84^\text{th}$ percentiles of the posterior to obtain a value of $\TIGM = 0.26_{-0.14}^{+0.25}$. For comparison, we also show the constraint on \TIGM{} at $z = 8 - 10$ ($\TIGM = 0.28_{-0.10}^{+0.15}$) obtained using \Lya{} measurements from multiple fields \citep[GOODS-N, GOODS-S, EGS, and Abell 2744, following][]{Tang2024_highz_lya} in Figure\ \ref{fig:T_IGM}.

Our measurement of \TIGM{} in EGS is fully consistent with the $z = 8 - 10$ IGM transmission constraint obtained from \Lya{} observations in multiple fields. This suggests that any ionized bubble that may be present in this field is no larger than the average at this redshift. However, if the average bubble radius at this redshift is $R_b \sim 2$\,pMpc or larger, this may still imply that such a bubble does exist in the EGS volume, which could contain all of the LAEs that have been observed. Thus, we next quantify whether our constraint is consistent with a bubble with the minimum 2\,pMpc radius necessary to contain the two bright LAEs in this volume by estimating what the IGM transmission would be for a galaxy in the center of a bubble with radius $R_b = 2$\,pMpc at $z = 8.7$.

Following \citet{Tang2024_highz_lya}, we model the transmission of \Lya{} from a galaxy with given absolute magnitude (\Muv) and systemic redshift ($z_s$) embedded inside a bubble with radius $R_b$ as
\begin{align} \label{eqn:Rb_TIGM}
\TIGM( & z_s, \Muv, x_\textsc{hi}, R_b) = \nonumber \\
& \int_{-\infty}^{\infty} \text{d}v \, J_{\Lya}(v) \exp[-\tau_\textsc{igm}(z_s, \Muv, x_\textsc{hi}, R_b, \lambda_\text{obs}(v))],
\end{align}
where $x_\textsc{hi}$ is the local neutral hydrogen fraction in the IGM near the bubble, $J_{\Lya}(v)$ is the normalized \Lya{} velocity profile emerging from the ISM and CGM of the galaxy before encountering the IGM, and $\tau_\textsc{igm}(z_s, \Muv, x_\textsc{hi}, R_b, \lambda_\text{obs}(v))$ is the optical depth provided by the IGM to the \Lya{} profile. For $J_{\Lya}(v)$, we normalize the composite \Lya{} line profile at $z \sim 5$ measured by \citet{Tang2024_z5_lya}, which is a single-peaked profile that peaks at a velocity offset of $\Delta v \sim 230$\,km\,s$^{-1}$, such that $\int_{-\infty}^{\infty} J_{\Lya}(v) \text{d}v = 1$. This assumes that any evolution in \Lya{} transmission at $z > 5$ is primarily due to evolution in the IGM. $\tau_\textsc{igm}$, the total optical depth provided by the IGM to \Lya, is the sum of two effects: resonant scattering due to residual intergalactic neutral hydrogen inside ionized bubbles infalling onto galaxies ($\tau_\text{infall}$) and damping wing attenuation in the neutral IGM outside the ionized bubble \citep[$\tau_\textsc{dw}$;][]{Miralda-Escude1998}.

We calculate the optical depth provided by infalling gas, $\tau_\text{infall}$, following the methods of \citet{Mason2018}. In this model, for a galaxy with given \Muv, gas from the IGM is assumed to be infalling at the circular velocity of the host halo of. Assuming a Navarro-Frenk-White \citep{Navarro1997} profile, this circular velocity is $v_c = [10 \text{G} M_h(\Muv) H(z_s)]^{1/3}$, where $G$ is the gravitational constant and $H(z_s)$ is the Hubble parameter at $z_s$. We assume that the infalling gas is infinitely optically thick to photons at velocities less than $v_c$ and is optically thin at velocities greater than $v_c$ (i.e. $\tau_\textsc{hii} = \infty$ at $v \leq v_c$ and $\tau_\textsc{hii} = 0$ at $v > v_c$). We estimate the halo mass, $M_h$, from the $\Muv - M_h$ relation at $z = 9$ from \citet{mason2015} (the closest redshift for which \citealt{mason2015} calculated an $\Muv - M_h$ relation to our redshift of interest, $z_s = 8.7$), which leads to brighter galaxies having larger halo masses, and therefore larger circular velocities. Thus, \Lya{} photons from bright galaxies are completely attenuated out to larger velocities on the red side of \Lya{} line center than from faint galaxies.

We calculate the optical depth provided by the damping wing in the IGM, $\tau_\textsc{dw}$, by integrating along the line of sight from $z_s$ to the end of reionization, $z_\text{reion}$. We evaluate $\tau_\textsc{dw}$ as a function of observed wavelength, $\lambda_\text{obs} = \lambda_\text{emit} (1 + z_s)$, corresponding to a \Lya{} photon emitted at a wavelength of $\lambda_\text{emit}$ from a galaxy at redshift $z_s$. Then, following the prescription of \citet{Mason2020}, the damping wing optical depth at $\lambda_\text{obs}$, $\tau_\textsc{dw}(\lambda_\text{obs})$, is
\begin{equation} \label{eqn:tau_dw}
\tau_\textsc{dw}(\lambda_\text{obs}) = \int_{z_s}^{z_\text{reion}} \text{d}z \, c \, \frac{\text{d}t}{\text{d}z} \, x_\textsc{hi}(z) \, n_\textsc{h}(z) \, \sigma_{\Lya}\Big(\frac{\lambda_\text{obs}}{1 + z}, T\Big),
\end{equation}
where $\text{d}t / \text{d}z = -1 / [(1 + z) \cdot H(z)]$ relates changes in linear time to changes in redshift, $x_\textsc{hi}(z)$ and $n_\textsc{h}(z)$ are the neutral hydrogen fraction and the proper volume density of hydrogen at the redshift under consideration, respectively, and $\sigma_{\Lya}(\lambda_\text{obs} / (1 + z), T)$ is the \Lya{} scattering cross-section in a gas with kinetic temperature $T$ at a wavelength of $\lambda_\text{obs} / (1 + z)$ (i.e. the observed wavelength of the emitted photon in the frame of the absorbing gas at redshift $z$, which accounts for the cosmological redshifting of the photon as it travels). The \Lya{} scattering cross-section is described well by a Voigt function \citep[for the full functional form, see equations 15 and 16 in][]{dijkstra2014}.

We note that the $z \sim 5$ \Lya{} profile that we assume as the intrinsic profile in this work is expected to already include the effects of infalling gas, as it was measured by \citet{Tang2024_z5_lya} after \Lya{} photons emerge from the galaxy into the IGM. That is, to first order, we expect $\tau_{\text{infall}, z = 5} = \tau_{\text{infall}, z=8.7}$. In this case, because we are interested in the relative increase in \Lya{} attenuation between $z = 5$ and $z = 8.7$  due to the neutral IGM, we can assume that $\tau_\textsc{igm} = \tau_\textsc{dw}$. However, the density of the IGM increases with increasing redshift and the average UV luminosity of our photometric sample is brighter than that of the sample used by \citet{Tang2024_z5_lya} to construct the $z \sim 5$ composite. Thus, $\tau_\text{infall}$ may be an important factor, though it likely is not infinite as presented above. Fully quantifying the effect of the redshift and luminosity dependence of $\tau_\text{infall}$ on \TIGM{} is beyond the scope of this work, but we present two limiting cases of \TIGM{} where we assume that $\tau_\textsc{igm} = \tau_\textsc{dw}$ and that $\tau_\textsc{igm} = \tau_\text{infall} + \tau_\textsc{dw}$, with $\tau_\text{infall}$ calculated as described previously.

Following \citet{Mason2020}, we evaluate Equation\ \eqref{eqn:tau_dw} in two steps. First, we integrate from $z_s$ to the redshift of the first neutral hydrogen patch that defines the edge of the bubble, $z_\text{bubble}$. Then, we integrate from $z_\text{bubble}$ to $z_\text{reion}$. We assume that the IGM is entirely ionized inside the bubble (i.e. $x_\textsc{hi} = 0$ from $z_s$ to $z_\text{bubble}$), so in practice, only the integral from $z_\text{bubble}$ to $z_\text{reion}$ is nonzero. We adopt a gas temperature of $T = 1000$\,K outside the bubble, broadly consistent with constraints on X-ray heating of the IGM from upper limits on the 21\,cm power spectrum during reionization \citep{HERACollaboration2023}. We fix $x_\textsc{hi}(z) = 1$, as the optical depth is dominated by attenuation in the first neutral hydrogen patch outside of the bubble \citep[e.g.][]{Mesinger2008, Mason2025}; \Lya{} photons are redshifted to increasingly long wavelengths as they travel such that the scattering cross-section decreases rapidly at redshifts increasingly far from $z_s$, so we expect attenuation in further neutral patches to be less than that in the first neutral region.

\begin{figure*}
    \includegraphics[width=\textwidth]{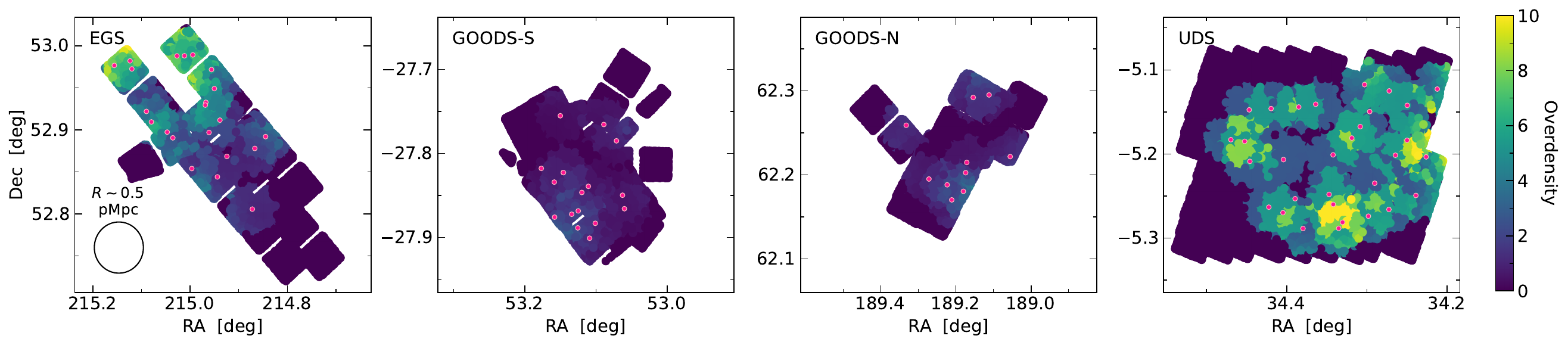}
    \caption{\label{fig:field_comparison}The on-sky positions (pink points) of the $z \sim 8.7$ galaxy candidates and the corresponding implied overdensities (colormap, where yellow corresponds to higher density regions and dark purple regions are lower density) in EGS, GOODS-S, GOODS-N, and UDS (from left to right). All images are on the same angular scale. Overdensities are calculated in circles with radii defined by the angular scale corresponding to 0.5\,pMpc at $z = 8.7$.}
\end{figure*}

We calculate both limiting cases of $\tau_\textsc{igm}$ and evaluate Equation\ \eqref{eqn:Rb_TIGM} for a galaxy at $z_s = 8.7$ with absolute UV magnitude of $\Muv = -20.7$ (similar to the median UV magnitude of the spectroscopic sample confirmed in EGS) that is located at the center of an $R_b = 2$\,pMpc bubble. We find that such a galaxy would be expected to have a \Lya{} transmission of $\TIGM = 0.63$ when $\tau_\textsc{igm} = \tau_\textsc{dw}$ and $\TIGM = 0.53$ when $\tau_\textsc{igm} = \tau_\text{infall} + \tau_\textsc{dw}$ (grey shaded region in Figure\ \ref{fig:T_IGM}). In comparison, 91\,per\,cent of the probability distribution for \TIGM{} that we have calculated in this field lies at $\TIGM < 0.63$ (86\,per\,cent at $\TIGM < 0.53$). In either case, we conclude that this implies that it is moderately unlikely (inconsistent at $\sim 1 - 1.5\sigma$) for this region to host a single ionized bubble with radius $R_b = 2$\,pMpc, though we cannot rule out a smaller bubble ($R_b \sim 0.5 - 1$\,pMpc) that is expected to be more typical at $z \sim 8.7$ \citep[e.g.][]{Lu2024_bubble_sizes}, and that may be consistent with the $R_b \sim 0.2 - 0.5$ bubble sizes inferred by \citet{Hayes2023} at $z = 8.7$. Furthermore, the \Lya{} emission observed in the galaxies in this field may be facilitated significantly by intrinsic properties of the galaxies themselves, consistent with the presence of high-ionization ($\geq 47$\,eV) rest-UV emission lines and strong \ion{[O}{iii]}+\ion{H}{$\beta$} emission. Further study of the \Lya{} emission of the $z = 8.7$ galaxy population in the EGS field, as well as a deeper understanding of their physical properties, will be necessary to better constrain the size of any bubble that may exist in this volume. Additionally, more comprehensive characterization of galaxy overdensities and LAEs at $z \sim 9$ will be crucial to place this region of the Universe in context and gain a more complete understanding of the overall ionization state of the IGM at $z \sim 9$.

\section{Towards A Large Sample of Ionized Bubbles at \texorpdfstring{$\lowercase{z} \sim 9$}{z=9}} \label{sec:discussion}

The simple fact that the only two $z \gtrsim 8$ LAEs known before \textit{JWST} \citep{Zitrin2015, Larson2022} were separated by only $\sim 4$\,pMpc prompted significant observational effort in the EGS field to characterize the volume, as it was potentially a unique site in which the middle stages of reionization could be studied. Such studies were challenging from the ground due to limited sensitivity that made it impossible to observe faint objects, as would be required to fully characterize the properties of any ionized bubble present. However, with \textit{JWST}, observing \Lya{} emission at $z \gtrsim 8$ along with any associated galaxy overdensities and ionized bubbles has become much more accessible.

In this work, we have found a mild photometric galaxy overdensity in the EGS field at $z \sim 9$ close to EGSY8p7, but current spectroscopic observations with \textit{JWST} do not find evidence of a very large, $R_b \gtrsim 2$\,pMpc ionized bubble. Given that galaxies in overdensities are thought to be producing copious amounts of ionizing photons that create large ionized bubbles \citep[e.g.][]{Dayal2018, Qin2022}, an overdensity without an associated ionized bubble would be unexpected, suggesting that there may be a smaller ionized region in the EGS volume. This may also be more consistent with theoretical expectations for the reionization topology at these redshifts \citep[e.g.][]{Lu2024_bubble_sizes}. However, observational constraints on ionized bubbles are still highly incomplete. The ionized bubble candidate at $z = 8.7$ in EGS is currently the most well characterized of any such bubble at $z > 7$, but current NIRCam imaging and spectroscopy in EGS is largely restricted to the CEERS imaging area that probes the vicinity of EGSY8p7. Thus, current data largely does not constrain the extent and morphology of any ionized region(s) that may extend towards the other bright LAE in the field. Additionally, this single sightline provides only one view of the reionization process. Extending the imaging and spectroscopy in EGS to probe a larger volume at $z = 8.7$, and studying galaxy overdensities and \Lya{} emission at $z \sim 9$ in other fields, will be essential to contextualize the bubble candidate in EGS and obtain a complete understanding of the early stages of the reionization process. Thus, we now begin this process by placing the EGS $z \sim 8.7$ overdensity and candidate bubble in context with other volumes probed by legacy extragalactic deep fields, primarily focusing on galaxy overdensities measured from photometric samples identified in the Great Observatories Origins Deep Survey-North and -South (GOODS-N and GOODS-S) and Ultra Deep Survey (UDS) field. Then, we make forecasts for future constraints on ionized bubble sizes that will be enabled by deep \textit{JWST} rest-UV spectroscopy.

\subsection{Galaxy overdensities as ionized bubble tracers}

\begin{figure*}
    \includegraphics[width=\textwidth]{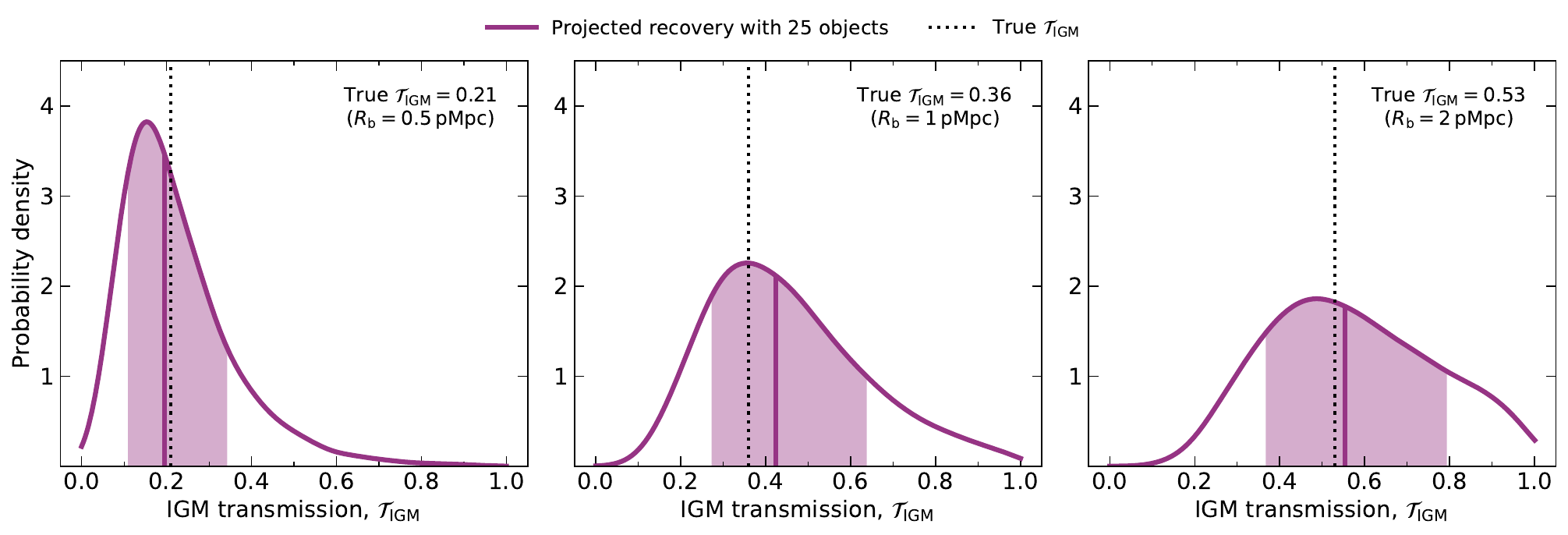}
    \caption{\label{fig:T_IGM_forecast}Examples of our forecasted constraints on the IGM transmission from a sample of 25 galaxies with spectroscopic observations at a $5\sigma$ line flux depth of $\sim 5.3 \times 10^{-19}$\,erg\,s$^{-1}$\,cm$^{-2}$. From left to right, we show the forecasts for three values of $\TIGM = 0.21, 0.36, 0.53$, approximately corresponding to ionized bubbles with radii of $R_b = 0.5, 1, 2$\,pMpc. We show our forecasts for the recovered constraint on \TIGM{} as the pink distributions and shaded regions (where the shaded regions correspond to the 68\,per\,cent credible intervals), and the input values of \TIGM{} as the dashed black lines. We recover the true value of \TIGM{} within the 68\,per\,cent credible interval for all three input values of \TIGM{} and the smallest value of \TIGM{} is recovered with the smallest uncertainties.}
\end{figure*}

To identify potential overdensities and ionized bubbles in other volumes, we reduce publicly available imaging data, then perform detection, photometry, and selection using the same methods as we used in EGS. We emphasize that we adopt the same color and photometric redshift selection criteria as we used for our EGS selection (Section\ \ref{sec:overdensity}) to ensure consistency across heterogeneous filter coverage in the various fields. We also require that all candidates are brighter than $m \leq 28.1$ to account for varying imaging depths (with the magnitude limit set by the depth of the field with the most shallow imaging). Performing this selection results in 22 candidates in EGS, 10 candidates in GOODS-N, 17 candidates in GOODS-S, and 31 candidates in UDS. We account for the selection functions in each field using the same methods as we used for EGS (Section\ \ref{subsec:overdensity_calc}), then quantify the overdensities by calculating observed surface densities and comparing to expected surface densities implied by measurements of the UV luminosity function \citep{Bouwens2021, Finkelstein2023, Donnan2024}. We note that we are interested in ionized bubbles and overdensities at $z \sim 9$ when the IGM is still likely highly neutral \citep[$\sim 70 - 90$\,per\,cent; e.g.][]{Tang2024_highz_lya}, implying that such structures would extend over physical scales smaller than the imaging footprints \citep[$R \sim 0.5$\,pMpc, corresponding to $R \sim 1.9$\,arcmin at $z = 8.7$;][]{Lu2024_bubble_sizes}. Thus, rather than calculating the average overdensity in each field, we create maps of overdensities over each field. Specifically, we calculate observed surface densities in circles with 1.9\,arcmin (0.5\,pMpc) radii projected on the sky at $z = 8.7$, then compare the spatially varying observed surface densities to the expected surface density. We show the resulting overdensity maps in Figure\ \ref{fig:field_comparison}. Of the four fields, we find that EGS and UDS have the largest overdensities on scales of $R \sim 0.5$\,pMpc, with overdensities by factors of $\sim 3 - 5$ spanning scales of a few arcminutes, and overdensities up to a factor of $\sim 10$ in smaller areas. We also identify a factor of $\sim 4 - 5$ overdensity in GOODS-N on a smaller scale than in EGS or UDS; otherwise, we find that the GOODS fields are of average density. These surface densities suggest that UDS may be the most likely candidate besides EGS to host ionized bubble(s) at $z \sim 9$, while the GOODS fields may largely be probing volumes that are still largely neutral at $z \sim 9$.

\subsection{Prospects for \texorpdfstring{\Lya}{Lya} constraints on ionized bubble sizes}

Quantifying the sizes of ionized bubbles that may exist in any of these fields will require deep spectroscopy in the rest-frame UV of a large number of objects. In this work, we have taken the first steps to compare our observed \Lya{} transmission to that expected from an ionized bubble with a given radius to provide insights into the bubble size. However, such measurements still suffer from small sample sizes that lead to large uncertainties, making it challenging to distinguish between ionized bubble radii. For a galaxy at $z = 8.7$ with an absolute UV magnitude of $\Muv = -20.7$ (i.e. similar properties as the sample we have presented in this work), a bubble with radius $R_b = 0.5$\,pMpc implies a \Lya{} transmission of $\TIGM = 0.21$, while a galaxy inside a bubble with radius $R_b = 1$\,pMpc is expected to have a \Lya{} transmission of $\TIGM = 0.36$. Thus, distinguishing between bubble radii of $R_b = 0.5$\,pMpc and $R_b = 1$\,pMpc -- either or both of which may be present at $z \sim 9$ \citep{Lu2024_bubble_sizes}  -- would require measurements of the \Lya{} transmission to a precision of approximately $\pm 0.15$. For comparison, in this work, we have measured the transmission of \Lya{} with a precision of $-0.14/ +0.25$, suggesting that constraining ionized bubble sizes using IGM transmission will soon be possible with deep \textit{JWST} \Lya{} observations.

To quantitatively examine the prospects for \textit{JWST} spectroscopy to constrain the transmission of \Lya{} through the IGM at the precision necessary to distinguish between ionized bubble sizes, we forecast the \TIGM{} constraints that we may expect from observations at a depth known to be attainable with \textit{JWST} (e.g. 30 hours in G140M, comparable to GO program 9214, PIs C. Mason and D. Stark). We draw \Lya{} EWs from the intrinsic \Lya{} EW distribution at $z \sim 5 - 6$ found by \citet{Tang2024_z5_lya} and draw UV continuum luminosities from the $z \sim 9$ UV luminosity function found by \citet{Donnan2024}, then calculate the intrinsic line fluxes that would be emitted. Next, we attenuate the intrinsic fluxes using the IGM transmissions that correspond to bubbles with radii of $R_b = 0.5, 1, \text{ and } 2$\,pMpc ($\TIGM = 0.21, 0.36, \text{ and } 0.53$, respectively, when including contributions to the optical depth from both infalling gas and damping wing attenuation in the IGM) and calculate the observed EWs and uncertainties assuming the $5\sigma$ line depth estimated for the G140M observations of GO 9214 ($\sim 5.3 \times 10^{-19}$\,erg\,s$^{-1}$\,cm$^{-2}$). From these mock data, we infer \TIGM{} using the same MCMC as we used for the real data in EGS (Section\ \ref{sec:lya_transmission}). We run 1000 realizations of this \TIGM{} recovery test, randomly sampling new values of intrinsic \Lya{} EW and \Muv{} for each realization. In Figure\ \ref{fig:T_IGM_forecast}, we show examples of the IGM transmission constraint forecasts if 25 objects have \Lya{} observations, but we test multiple sample sizes ranging between $N = 5$ and $N = 50$ in steps of five and discuss the constraints on \TIGM{} as a function of the number of objects observed. We emphasize that these constraints are not designed to specifically predict the outcomes of GO 9214, but we have adopted the observational parameters of a real \textit{JWST} spectroscopic survey to ensure that our forecast accurately represents the observational capabilities of \textit{JWST}.

Overall, we find that our inference for \TIGM{} generally recovers the true value of \TIGM{} successfully and without significant bias (the true value of \TIGM{} falls within the 68\,per\,cent credible interval of the recovered \TIGM{} distribution for $\geq 70$\,per\,cent of our realizations). As expected, the uncertainties on the recovered value of \TIGM{} become smaller as the number of objects that have \Lya{} constraints (both detections and upper limits) increases. Additionally, as seen in Figure\ \ref{fig:T_IGM_forecast}, the recovered constraint on \TIGM{} is more precise for smaller values of true \TIGM{} due to the likelihood function (the \TIGM-dependent \Lya{} EW distribution) becoming more sharply peaked, and therefore more sensitive to small variations in \TIGM, as \TIGM{} decreases. For the smallest value of \TIGM{} that we test, $\TIGM = 0.21$ (corresponding to a bubble radius of $R_b = 0.5$\,pMpc), we find that a spectroscopic sample of 25 objects is sufficient to constrain \TIGM{} to $\TIGM = 0.21_{-0.10}^{+0.15}$, the minimum precision necessary to distinguish between $\TIGM = 0.21$ ($R_b = 0.5$\,pMpc) and $\TIGM = 0.36$ ($R_b = 1$\,pMpc). For larger values of $\TIGM = 0.36$ and $\TIGM = 0.53$, we find that samples of $\sim 40$ and $\sim 50$ objects with \Lya{} spectroscopy are required to reach similar precision. While these sample sizes are reasonably large, existing spectroscopic observations have already placed \Lya{} constraints on $10 - 20$ objects per field at $z = 8 - 10$ in the EGS and GOODS fields \citep[e.g.][]{Tang2024_highz_lya}, suggesting that the requisite sample sizes will be in reach of future \Lya{} follow up of photometric candidates with \textit{JWST}. We also note that with these large sample sizes, the uncertainties introduced by aperture effects of the NIRSpec MSA (Section\ \ref{sec:lya_transmission}) may be reduced, as the effects are significantly more likely to average out over a sample of $\geq 20$ objects than the four we present in this work.

\medskip

In this work, we have taken the first steps towards using \Lya{} emission from galaxies to constrain the reionization topology. In the near future, new theoretical methods that leverage the spatial distribution of \Lya{} transmission from galaxies within ionized bubbles \citep{Lu2024_bubble_mapping, Nikolic2025} will become increasingly informative. The transmission of \Lya{} is expected to be highest when the line-of-sight distance to the nearest patch of neutral hydrogen is longest, and thus is expected to vary spatially within a given ionized bubble. These spatial variations of \Lya{} transmission can be used to empirically detect the edges of ionized bubbles \citep{Lu2024_bubble_mapping}, where edges are identified by a decline in the transmission. Additionally, forward models of the spatial distribution of observed \Lya{} properties as a function of bubble position and size can be developed \citep{Nikolic2025}, which can then be compared with observations to infer ionized bubble properties. As the number of objects with \Lya{} constraints continue to grow, applying these methods will enable robust constraints on ionized bubble sizes that will significantly increase our understanding of the reionization process.

\section{Summary} \label{sec:summary}

In this work, we have presented rest-frame UV and optical NIRSpec observations targeting \Lya{} emission of galaxies tracing a possible ionized bubble at $z = 8.7$ in the EGS field, taken as part of \textit{JWST} GO program 4287. We have characterized the $z \sim 8.7$ galaxy population in the field and measured the transmission of \Lya{} implied by these observations. We have then examined the implications of both of these measurements for the presence of a very large ($R_b \gtrsim 2$\,pMpc) ionized bubble in the $z \sim 8.7$ volume probed by EGS. We summarize our key conclusions below.

\begin{enumerate}
    \item We have spectroscopically observed four galaxies at systemic redshifts of $\zsys = 8.7 \pm 0.02$ with physical separations ranging from 0.05\,pMpc to 1.9\,pMpc. If there is an ionized bubble large enough to contain the two bright LAEs in the volume ($R_b \gtrsim 2$\,pMpc), all of these galaxies could lie within it.
    \item We visually identify \Lya{} emission from two of the four galaxies (EGSY8p7 with $\text{EW}_{0, \Lya} = 7.6 \pm 2.2$\,\AA, CEERS-1025 with $\text{EW}_{0, \Lya} = 3.5 \pm 2.2$\,\AA) and place $3\sigma$ upper limits on the \Lya{} flux and EW of the remaining two objects (70289 with $\text{EW}_{0, \Lya} < 16.9$\,\AA, 89540 with $\text{EW}_{0, \Lya} < 28.9$\,\AA). The two LAEs have very strong \ion{[O}{iii]}+\ion{H}{$\beta$} ($\text{EW}_{0, \ion{[O}{iii]}+\ion{H}{$\beta$}} \gtrsim 1200$\,\AA), with EWs among the most extreme known in $z \sim 9$ galaxies, suggesting that these galaxies have very hard ionizing radiation fields that may facilitate the production of intrinsically strong \Lya. In contrast, the two galaxies not emitting \Lya{} have significantly weaker \ion{[O}{iii]}+\ion{H}{$\beta$} emission ($\EW_{0, \ion{[O}{iii]}+\ion{H}{$\beta$}} \sim 600-700$\,\AA), hinting at weaker ionizing radiation fields.
    \item We find a factor of $\sim 2.4 - 3.6$ overdensity within the 5\,arcmin ($\sim 1.4$\,pMpc in projection at $z = 8.7$) radius of EGSY8p7 (an update to the original measurement by \citealt{Whitler2024}), hinting at the presence of a galaxy population that may be able to contribute significant amounts of ionizing flux towards creating a large ionized bubble.
    \item We combine our new measurements of \Lya{} in four galaxies with \Lya{} measurements and upper limits reported by \citet{Tang2024_highz_lya} at redshifts of $z = 8.6 - 8.8$ (spanning a line-of-sight separation of $\sim 5.3$\,pMpc) to infer the transmission of \Lya{} through the IGM, \TIGM, in this volume. We find $\TIGM = 0.26_{-0.14}^{+0.25}$, which is fully consistent with the average IGM transmission of $\TIGM = 0.28_{-0.10}^{+0.15}$ in four independent fields at $z = 8 - 10$ measured by \citet{Tang2024_highz_lya}. This transmission is in mild tension with the IGM transmission of $\TIGM \sim 0.53 - 0.63$ that would be expected inside an ionized bubble with radius $R_b = 2$\,pMpc, implying that it is moderately unlikely for an ionized bubble of this size to exist in this volume \citep[consistent with theoretical expectations for the reionization topology at $z \sim 9$; e.g.][]{Lu2024_bubble_sizes}.
    \item We search other large extragalactic imaging datasets for overdensities that may hint at ionized bubbles. We find that though EGS does not have evidence of an extremely large, $R_b \gtrsim 2$\,pMpc ionized bubble, it is still more overdense on the projected scales of ionized bubbles expected at $z \sim 8.7$ than two of the three other fields we investigate. This may imply that an ionized bubble with a radius $< 2$\,pMpc may still be present.
    \item Confirming the presence of a smaller bubble will require deep \Lya{} spectroscopy, so we forecast the constraints on the IGM transmission that will be accessible with future \textit{JWST} \Lya{} observations. We find that distinguishing between a particularly large $R_b \sim 2$\,pMpc bubble and more typical, $R_b \sim 1$\,pMpc or $R_b \sim 0.5$\,pMpc bubbles using \Lya{} transmission measurements will be possible with sample sizes of $\gtrsim 25$ objects with observations at a depth comparable to GO program 9214 (30 hours in G140M, corresponding to an estimated $5\sigma$ line flux depth of $\sim 5.3 \times 10^{-19}$\,erg\,s$^{-1}$\,cm$^{-2}$). We emphasize that new methods that take advantage of the spatial variations of \Lya{} transmission \citep{Lu2024_bubble_mapping, Nikolic2025} will be able to extend this measurement of the average \Lya{} transmission to robustly map ionized bubbles and constrain the reionization topology.
\end{enumerate}

\section*{Acknowledgements}

We thank the anonymous referee for their constructive comments that helped to improve this paper. LW acknowledges support from the Gavin Boyle Fellowship at the Kavli Institute for Cosmology, Cambridge and from the Kavli Foundation. DPS acknowledges support by the National Science Foundation under Grant No. AST-2109066. CAM acknowledges support by the European Union ERC grant RISES (101163035), Carlsberg Foundation (CF22-1322), and VILLUM FONDEN (37459). Views and opinions expressed are those of the author(s) only and do not necessarily reflect those of the European Union or the European Research Council. Neither the European Union nor the granting authority can be held responsible for them. ZC, TYL, GPL and AH acknowledge support by VILLUM FONDEN (37459). The Cosmic Dawn Center (DAWN) is funded by the Danish National Research Foundation under grant DNRF140.

This work is based on observations made with the NASA/ESA/CSA JWST. The data were obtained from the Mikulski Archive for Space Telescopes (MAST) at the Space Telescope Science Institute, which is operated by the Association of Universities for Research in Astronomy, Inc., under NASA contract NAS 5-03127 for JWST. These observations are associated with program 4287, support for which was provided by NASA through a grant from the Space Telescope Science Institute, which is operated by the Association of Universities for Research in Astronomy, Inc., under NASA contract NAS 5-03127. The authors acknowledge the CEERS team led by Steven Finkelstein and the RUBIES team led by Anna de Graaff and Gabriel Brammer for developing their observing programs with zero-exclusive-access periods. The data used in this work can be found on MAST at the following DOI: \href{https://doi.org/10.17909/19rp-r511}{10.17909/19rp-r511}.

This project made use of High Performance Computing (HPC) resources supported by the University of Arizona TRIF, UITS, and Research, Innovation, and Im pact (RII) and maintained by the UArizona Research Technologies department. We respectfully acknowledge the University of Arizona is on the land and territories of Indigenous peoples. Today, Arizona is home to 22 federally recognized tribes, with Tucson being home to the O’odham and the Yaqui. Committed to diversity and inclusion, the University strives to build sustainable relationships with sovereign Native Nations and Indigenous communities through education offerings, partnerships, and community service.

This work made use of the following software: \texttt{numpy} \citep{Harris2020}; \texttt{scipy} \citep{2020SciPy-NMeth}; \texttt{astropy}\footnote{\url{http://www.astropy.org}}, a community-developed core Python package and an ecosystem of tools and resources for astronomy \citep{AstropyCollaboration2013, AstropyCollaboration2018, AstropyCollaboration2022}; \texttt{pandas} \citep{the_pandas_development_team_2024_13819579}; \texttt{matplotlib} \citep{Hunter2007}; \texttt{shapely} \citep{gillies_2024_shapely}; \texttt{Source Extractor} \citep{Bertin1996} via \texttt{sep} \citep{Barbary2016}; \texttt{photutils}, an Astropy package for
detection and photometry of astronomical sources \citep{larry_bradley_2025_14889440}; \texttt{BEAGLE} \cite{Chevallard2016}; \texttt{multinest} \citep{Feroz2008, Feroz2009, Feroz2019}; \texttt{emcee} \citep{Foreman-Mackey2013}; \texttt{sedpy} \citep{johnson_2021_4582723}

\section*{Data Availability}

The \textit{JWST}/NIRSpec observations from \textit{JWST} GO program 4287, the ancillary RUBIES (GO program 4233) and CEERS (ERS program 1345) spectra, and the NIRCam imaging from CEERS and GO program 2234 are all available through the Mikulski Archive for Space Telescopes (\url{https://mast.stsci.edu/}) under their respective proposal IDs. Reduced data and analysis products will be made available upon reasonable request to the corresponding author.



\bibliographystyle{mnras}
\bibliography{refs} 







\bsp	
\label{lastpage}
\end{document}